\documentclass[onecolumn,a4paper,accepted=2021-04-12]{quantumarticle}
\pdfoutput=1
\usepackage{algorithm,algcompatible,amsmath,graphicx,typearea,here,color,amsthm,docmute,grffile}
\usepackage[numbers,sort&compress]{natbib}
\usepackage[subrefformat=parens]{subcaption}
\typearea{16}
\algnewcommand\INPUT{\item[\textbf{Input:}]}%
\algnewcommand\OUTPUT{\item[\textbf{Output:}]}%
\algnewcommand\RETURN{\item[\textbf{return}]}%
\newcommand{\f}[2]{\mathcal{F}^{(#1)}(#2)}
\newcommand{\fhaar}[1]{\mathcal{F}^{(#1)}_{\rm Haar}(N)}
\newcommand{\fhaarn}[1]{\mathcal{F}^{(#1)}_{\rm Haar}(2^n)}
\newcommand{\expr}[0]{\mathcal{E}}

\newcommand{\heabase}[2]{C_{\rm HEA}^{#1,#2}}
\newcommand{\altbase}[3]{C_{\rm ALT}^{#1,#2,#3}}
\newcommand{\tenbase}[3]{C_{\rm TEN}^{#1,#2,#3}}
\newcommand{\heacustom}[1]{\heabase{#1}{n}}
\newcommand{\altcustom}[1]{\altbase{#1}{m}{n}}
\newcommand{\tencustom}[1]{\tenbase{#1}{m}{n}}
\newcommand{\hea}[0]{\heacustom{\ell}}
\newcommand{\alt}[0]{\altcustom{\ell}}
\newcommand{\ten}[0]{\tencustom{\ell}}
\newcommand{\fheabase}[3]{\f{#1}{\heabase{#2}{#3}}}
\newcommand{\faltbase}[4]{\f{#1}{\altbase{#2}{#3}{#4}}}

\newcommand{\fheacustom}[2]{\f{#1}{\heacustom{#2}}}
\newcommand{\faltcustom}[2]{\f{#1}{\altcustom{#2}}}
\newcommand{\ftencustom}[2]{\f{#1}{\tencustom{#2}}}
\newcommand{\fhea}[1]{\f{#1}{\hea}}

\newcommand{\ften}[1]{\f{#1}{\ten}}
\newcommand{\pf}[1]{P(#1, F)}
\newcommand{\phaar}[0]{P_{\rm Haar}(F)}
\newcommand{\ualayer}[1]{U_a{\left(#1\right)}}
\newcommand{\ublayer}[1]{U_b{\left(#1\right)}}
\newcommand{\uablock}[2]{U_a{\left(#1, #2\right)}}
\newcommand{\ubblock}[2]{U_b{\left(#1, #2\right)}}

\newcommand{\intone}[0]{\int_{\rm 1design}}
\newcommand{\inttwo}[0]{\int_{\rm 2design}}
\newcommand{\sumak}[1]{\sum_{k_{#1}^a=1}^4}
\newcommand{\sumbk}[1]{\sum_{k_{#1}^b=1}^4}

\newcommand{\lambdaa}[1]{\lambda_{k_{#1}^a}^{(m)}}
\newcommand{\lambdab}[1]{\lambda_{k_{#1}^b}^{(m)}}
\newcommand{\delthree}[1]{\Delta^{(k_{1#1}^a, k_{3#1}^a, k_{1#1}^b)}\left(\uablock{2}{#1}, \ubblock{2}{#1}\right)}
\newcommand{\delthreebase}[2]{\Delta^{(k_{1#1}^a, k_{3#1}^a, k_{1#1}^b)}\left(\uablock{2}{#2}, \ubblock{2}{#2}\right)}
\newcommand{\delthreeplus}[1]{\Delta^{(k_{1#1}^a, k_{3#1}^a, k_{1#1}^b)}\left(\uablock{2}{#1+1}, \ubblock{2}{#1+1}\right)}
\newcommand{\delsix}[2]{\Delta^{\left(k_{1#1}^a, k_{3#1}^a, k_{1#1}^b, k_{1#2}^a, k_{3#2}^a, k_{1#2}^b\right)}(\uablock{2}{#2}, \ubblock{2}{#2})}
\newcommand{\athreem}[0]{{\bf a}(3, m)}
\newcommand{\atwom}[0]{{\bf a}(2, m)}
\newcommand{\aellm}[0]{{\bf a}(\ell, m)}
\newcommand{\vzero}[0]{{\bf v}_0}
\newcommand{\vone}[0]{{\bf v}_1}
\newcommand{\vzeroi}[0]{{\bf v}_{0i}}
\newcommand{\vonei}[0]{{\bf v}_{1i}}
\newcommand{\vvec}[1]{{\bf v}_{#1}}
\newcommand{\gxd}[2]{g^{#1}_{#2}(X, D)}
\newcommand{\gxddef}[0]{g^{k}_{\alpha}(X, D)}
\newcommand{\CenterRow}[2]{
  \dimen0=\ht\strutbox%
  \advance\dimen0\dp\strutbox%
  \multiply\dimen0 by#1%
  \divide\dimen0 by2%
  \advance\dimen0 by-.5\normalbaselineskip%
  \raisebox{-\dimen0}[0pt][0pt]{#2}}

\newtheorem{th.}{Theorem}
\newtheorem{corollary}{Corollary}

\newcommand{\fig}[4]{
\begin{figure}
\centering
\includegraphics[width=160mm]{#1}
\vspace*{#4cm}
\caption{#2}
\label{#3}
\end{figure}
}

\newcommand{\figmedium}[3]{
\begin{figure}
 \centering
 \includegraphics[width=130mm,bb=0 0 1536 1200]{#1}
\caption{#2}
\label{#3}
\end{figure}
}

\begin{document}

\title{Expressibility of the alternating layered ansatz for quantum computation}
\author{Kouhei Nakaji and Naoki Yamamoto}
\affil{Department of Applied Physics and Physico-Informatics \& Quantum Computing Center,
Keio University, Hiyoshi 3-14-1, Kohoku, Yokohama, 223-8522, Japan}

\maketitle

\abstract{
The hybrid quantum-classical algorithm is actively examined as a technique applicable even 
to intermediate-scale quantum computers. 
To execute this algorithm, the hardware efficient ansatz is often used, thanks to its 
implementability and expressibility; 
however, this ansatz has a critical issue in its trainability in the sense that it generically suffers 
from the so-called gradient vanishing problem. 
This issue can be resolved by limiting the circuit to the class of shallow alternating layered 
ansatz. 
However, even though the high trainability of this ansatz is proved, it is still unclear whether 
it has rich expressibility in state generation. 
In this paper, with a proper definition of the expressibility found in the literature, we show 
that the shallow alternating layered ansatz has almost the same level of expressibility as 
that of hardware efficient ansatz. 
Hence the expressibility and the trainability can coexist, giving a new designing method for 
quantum circuits in the intermediate-scale quantum computing era. 
}


\section{Introduction}

Recent rapid progress in the hardwares for quantum computing stimulates researchers 
to develop new techniques that utilize even noisy intermediate-scale quantum (NISQ) 
\cite{NISQ} device for real applications such as machine learning and quantum chemistry. 
Especially, several hybrid quantum-classical algorithms have been actively examined, as 
a means to reduce the computational cost required on quantum computing part. 
The variational quantum eigensolver (VQE) \cite{vqe}, which trains the parameterized 
quantum circuit via a classical computation to decrease a given cost function, is typically 
considered as such a hybrid algorithm.

The critical point in this strategy is in the difficulty to design a suitable and implementable 
circuit ansatz that, after the training process, may produce an exact or well-approximating 
solution of a given problem. 
The hardware efficient ansatz (HEA) \cite{hardware-efficient-ansatz} is often used mainly 
because of the implementability on a hardware; 
this is a relatively shallow circuit ansatz, whose parameters are embedded in the angles of 
single-qubit rotation gates. 
However, it was proven in \cite{plateau} that, when those parameters are randomly chosen, 
the gradient of a standard cost function vanishes, meaning that the update of the cost 
function often gets stuck in the learning process before reaching the minimum.

For solving this vanishing gradient problem, several approaches have been proposed. 
Reference \cite{plateau-initialization} showed numerically that, with a special type of 
initializing method of the HEA, the vanishing gradient problem does not occur in an ansatz 
with ${\cal O}(1)$ qubits. 
Also, a quantum analogue of the natural gradient have been proposed in \cite{Carleo2019,Yamamoto2019}, the original classical version of which is often used to avoid similar 
vanishing gradient problems in neural networks. 
In this paper, we focus on the third approach given by Ref.~\cite{local-cost-function} that 
provides a method for devising a specific structure of the HEA ansatz, called the Alternating 
Layered Ansatz (ALT), which in fact provably does not suffer from the vanishing gradient 
problem. 
By definition, the class of ALT is included in that of HEA; the difference is that, while a HEA 
consists of multiple layers of single-qubit rotation gates and entanglers that in principle 
combines {\it all} qubits in each layer, the entangling gates contained in an ALT is restricted 
to entangle only {\it local} qubits in each layer. 
With this setting, the authors in \cite{local-cost-function} derived a strict lower bound of 
the variance of the gradient for an ALT with its parameters randomly chosen (more precisely, 
the ensemble of unitary matrices corresponding to each circuit block is 2-design) under the 
condition that the cost function is local (that is, the cost function is composed of local 
functions of only a small number of local qubits). 
By using this lower bound, it was also shown that the vanishing gradient problem can be 
resolved if the number of layers is of the order $O({\rm poly}(\log n))$ where $n$ is the 
total number of qubits, or roughly speaking if the circuit is shallow.

Then an important question arises; does ALT have a sufficient expressive power (expressibility) 
for generating a rich class of states, which contains the optimal or a well-approximating state? 
Because the set of ALTs is a subclass of that of HEA, one might argue that the expressibility of 
ALT could be much lower than that of the HEA; if this is the case, the ALT circuit may not 
generate a desired state even though the learning process is smoothly running. 
Thus, it is worth examining the expressibility of ALT in order to assess the practicality of this 
ansatz in executing the hybrid quantum-classical algorithm. 
In this paper, we study this problem using the expressibility measure introduced in 
\cite{expressibility} and show that, fortunately, the class of shallow ALTs has the same level 
of expressibility as that of HEAs. 
Therefore, the expressibility and the trainability of a quantum circuit can coexist, which means 
that the existing HEA found in the literature can be basically replaced with a simpler ALT 
without degradation of expressibility while acquiring a better trainability. 
That is, the ALT might be taken as a new standard ansatz in NISQ computing era.

The structure of this paper is as follows. 
Section~\ref{preliminary} is the preliminary, giving the definition of expressibility and the 
ansatzes. 
In Sec.~\ref{evaluation}, we show both theoretically and numerically that the expressibility 
of ALT is as high as that of HEA. 
In Sec.~\ref{vqe-section}, we show how much the value of expressibility is reflected to 
the result of VQE. 
Finally, we conclude with some remarks in Sec.~\ref{conclusion}.


\section{Preliminaries}
\label{preliminary}

In this section, we define indicators of the expressibility and introduce some circuit 
ansatzes.


\subsection{Indicators of the expressibility}

Following Ref.~\cite{expressibility}, we define the expressibility of a given circuit, by the 
randomness of states generated from the circuit, in terms of the frame potential and 
the Kullback-Leibler (KL) divergence.

\subsubsection*{Frame Potential}

To define the expressibility of a given circuit ansatz $C$, let us consider the deviation of 
the state distribution generated by $C$ from the Haar distribution, as follows; 
\[
     {\mathcal A}^{(t)}(C) = 
                \left\|
                \int_{\rm Haar} (|\psi\rangle\langle\psi|)^{\otimes t} d\psi 
                  - \int_{\Theta} (|\psi_{\theta}\rangle\langle\psi_{\theta}|)^{\otimes t} d\theta 
            \right\|_{HS},  
\]
where $\int_{\rm Haar}$ denotes the integration over the state $|\psi\rangle$ distributed with 
respect to the Haar measure, and $\| \cdot \|_{HS}$ is the Hilbert Schmidt distance. 
Also, $|\psi_{\theta}\rangle$ is the state generated by the ansatz $C$ characterized by the 
parameter $\theta\in\Theta$, e.g., $|\psi_{\theta}\rangle=U_C(\theta)|0\rangle, \theta\in\Theta$, 
where $U_C(\theta)$ is the unitary operator corresponding to $C$ and $|0\rangle$ is an 
initial state. 
Then we call that the ansatz $C$ with smaller ${\mathcal A}^{(t)}(C)$ has a higher 
expressibility. 
This definition is justified by the following reason. 
That is, because the state $|\psi\rangle$ generated from the Haar distribution can in principle 
represent an arbitrary state, the condition ${\mathcal A}^{(t)}(C)\approx 0$ implies that the 
ansatz $C$ can generate almost all states possibly including the optimal solution (e.g., the ground 
state in VQE); also, in this case the states generated from $C$ are almost equally distributed, 
which is particularly favorable if little is known about the problem.

To compute ${\mathcal A}^{(t)}(C)$, we instead focus on the following $t$-th generalized frame potential 
\cite{frame-potential-2} of $C$: 
\begin{flalign}
\label{frame-potential}
    \f{t}{C} = \int_{\Phi} \int_{\Theta} |\langle \psi_{\phi}|\psi_{\theta}\rangle|^{2t} d\phi d\theta, 
\end{flalign}
where both $\Theta$ and $\Phi$ represent the same set of parameters of $C$. 
In the present paper, we simply call it the $t$-th frame potential. 
In particular, the $t$-th frame potential of $N$-dimensional states distributed with respect 
to the Haar measure, is given by 
$\fhaar{t} = \int_{\rm Haar}\int_{\rm Haar}|\langle\psi|\psi^{\prime}\rangle|^{2t}
d\psi d\psi^{\prime}$. 
The point of introducing the frame potential is because these quantities are linked to 
${\mathcal A}^{(t)}(C)$ in the following form; that is, for an arbitrary positive integer $t$, 
it holds
\begin{flalign}
\label{frame-potential ineq}
     \f{t}{C} - \fhaar{t} 
          = {\mathcal A}^{(t)}(C) \geq 0. 
\end{flalign}
The equality in the last inequality holds if and only if the ensemble of $|\psi_{\theta}\rangle$ 
is a state $t$-design 
\cite{frame-potential-t-design-1,frame-potential-t-design-2,frame-potential-t-design-3}. 
Thus, the ansatz $C$ with smaller $\f{t}{C}$ has a higher expressibility. 
Also $\f{t}{C}$ is lower bounded by $\fhaar{t}$, meaning that the frame potential can be 
used as an indicator for quantifying the non-uniformity in the state distribution. 
In Sec.~\ref{moment}, we calculate $\f{1}{C}$ and $\f{2}{C}$ for several ansatz $C$.

\subsubsection*{KL-Divergence}

Note that the frame potential \eqref{frame-potential} is the $t$-th moment of the fidelity 
$F = |\langle \psi_{\theta}|\psi_{\phi}\rangle|^2$, where the circuit parameters 
$\theta\in\Theta$ and $\phi\in\Phi$ are randomly sampled from the circuit ansatz $C$. 
Hence the probability distribution of $F$, denoted by $\pf{C}$, contains more information 
for quantifying the randomness of $C$ than $\f{t}{C}$, and thereby the following measure 
can be used to quantify the expressibility of $C$:
\begin{flalign}
       \operatorname{\expr(C)}=D_{\mathrm{KL}}\left(\pf{C} \| P_{\mathrm{Haar}}(F)\right)
        = \int_0^1 \pf{C} \log \frac{\pf{C}}{P_{\mathrm{Haar}}(F)} dF, 
\end{flalign}
where $D_{KL} (q \| p) $ is the KL divergence between $q$ and $p$. 
Also $P_{\rm Haar}(F)$ is the probability distribution of the fidelity 
$F=|\langle\psi|\psi^{\prime}\rangle|^2$, where $|\psi\rangle$ and $|\psi^{\prime}\rangle$ 
are sampled according to the Haar measure. 
In Ref.~\cite{average-fidelity}, $ P_{\mathrm {Haar}} (F) = (N-1) (1-F)^{N-2}$ is derived, 
where $ N $ is the dimension of Hilbert space. 
Because in general $D_{KL} (q \| p) =0$ iff $q=p$, the anzatz $C$ with smaller value of 
$\expr(C)$ has a higher expressibility. 
Thus, $\expr(C)$ can also be used an indicator for quantifying the non-uniformity of an ansatz.

Lastly recall that the $t$-th moment of $\pf{C}$ and $P_{\rm Haar}(F)$ are $\f{t}{C}$ and 
$\fhaar{t}$, respectively. 
Thus, if the values of $\f{t}{C}$ is close to $\fhaar{t}$, the value of $\expr(C)$ is close to zero.


\subsection{Ansatzes}
\label{ansatzes}

Here we describe the three types of ansatzes investigated in this paper. 

\subsubsection*{Hardware Efficient Ansatz}

The HEA circuit consists of multiple layers of parametrized single qubit gates and entanglers 
which entangle all qubits. 
In the following, let $\hea$ be the class of HEA with $\ell$ layers where each layer contains 
$n$-qubits. 

\subsubsection*{Alternating Layered Ansatz}

The ALT introduced in \cite{local-cost-function} also consists of multiple layers but the 
components of each layer are different from those of the HEA as follows. 
That is, each layer has some separated blocks, where each block has parametrized 
single-qubit rotation gates and fixed entanglers that entangle all qubits inside the block. 
The probability distributions of those angle parameters are independent in all blocks and 
all layers. 
In this paper, we further restrict the class of ALT as follows. 
First, as in HEA, the entire circuit is composed of $\ell$ layers, where each layer contains 
$n$ qubits. 
Then, we assume that, in the odd-number-labeled layers, each block contains $m$ qubits, 
so that $m$ is an even number and $n/m$ is an integer. 
In other words, the odd-number-labeled layers contain $n/m$ blocks which operate on 
$\{1, \ldots, m\}$, $\{m+1\ldots, 2m\}$, ..., and $\{n-m + 1, \ldots, n\}$ qubits. 
As for the even-number-labeled layers, they contain $n/m + 1$ blocks which operate on 
$\{1,\ldots, m/2\}$, $\{m/2 + 1,\ldots, 3m/2\}$, ..., and $\{n-m/2 + 1, \ldots, n\}$ qubits; 
that is, the first and the last block operate on $m/2$ qubits, while the others operate on 
$m$ qubits. 
In the following, we use $\alt$ to denote the class of ALT with the above-defined indices.

In \cite{local-cost-function}, it is proved that the vanishing gradient problem can be avoided 
if the following conditions are satisfied; 
(i) each term of the cost Hamiltonian, $H=\sum_k H_k$ is local, meaning that each term $H_k$ 
is composed of less than $m$ neighboring qubits, 
(ii) the ensemble of unitary matrices in each block is 2-design, and 
(iii) the number of layers, $\ell$, is of the order $O({\rm poly}(\log n))$. 
Note that each block needs to be deep enough for the distribution of corresponding unitary 
matrices to be close to 2-design.

\subsubsection*{Tensor Product Ansatz} 
In addition to HEA and ALT, we here introduce the class of tensor product ansatz (TEN) as a 
relatively weak ansatz. 
This ansatz also consists of $\ell$ layers, where each layer contains $n$ qubits, and each layer 
contains $n/m$ blocks ($n/m$ is assumed to be an integer), which contain single-qubit 
rotation gates and entanglers combining all qubits in the block. 
Throughout all the layers, the blocks operate on $\{1, \ldots, m\}$, $\{m+1\ldots, 2m\}$, ..., and 
$\{n-m + 1, \ldots, n\}$ qubits. 
Thus, TEN always generates a product state of the form 
$|\psi_1\rangle \otimes \cdots \otimes |\psi_{n/m}\rangle$ where each state is composed of 
$m$ qubits. 
Let $\ten$ be the class of TEN described above. 

\subsubsection*{} 
\vspace{-0.7cm}
In Fig.~\ref{fig-ansatzes}, we show examples of the structures of these ansatzes.

\figmedium{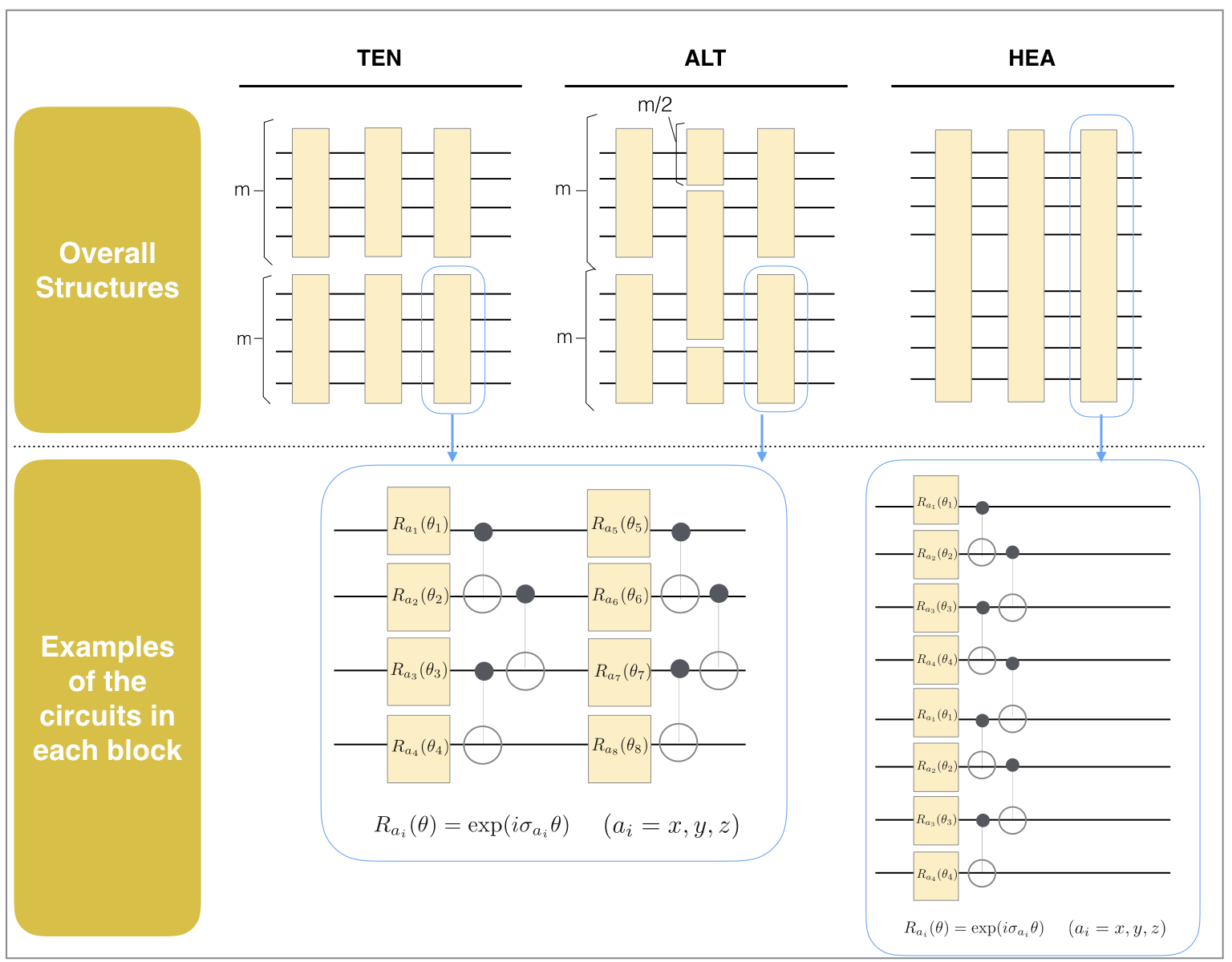}
{Examples of TEN, ALT, and HEA introduced in Section~\ref{ansatzes}. 
The upper figures are the overall structures of the ansatzes in the case $n=8$ and 
$\ell=3$ (and $m=4$ for TEN and ALT).
The circuits inside the blocks are exemplified in the lower figure, where $\sigma_{a_i}$ 
($a_i \in \{x, y, z\}$) is the Pauli matrix operating on a qubit and $\theta_i \in [0, 2\pi]$ 
is the parameter. 
}
{fig-ansatzes}


\section{Expressibility of the circuit ansatzes}
\label{evaluation}

In this section, we give some analytical expressions as well as upper bounds of the first 
and the second frame potentials of the three ansatzes introduced in Section \ref{ansatzes}, 
showing that the shallow ALT has almost the same expressibility as that of HEA. 
This result will be further confirmed by a numerical simulation in terms of the KL-divergence.


\subsection{Analytical expression of the frame potential of the ansatzes}
\label{moment}

First of all, to compute the frame potentials of each ansatz, we assume that the ensemble 
of the unitary matrices corresponding to $\hea$ is 2-design. 
Similarly, the ensemble of the unitary matrices corresponding to each block of $\alt$ and 
$\ten$ are assumed to be 2-design. These assumptions are also adopted in the discussion 
of \cite{plateau} and \cite{local-cost-function}. 
Such ensemble of unitary matrices can be generated by randomly choosing the parameters 
of the circuit having a specific structure.

Before going into the detail, we show some integration formulae for random unitary 
matrices \cite{haar-integral}. 
First, if the ensemble of $n\times n$ unitary matrices $\{ U \}$ is 1-design, the following 
formula holds: 
\begin{flalign}
\label{one-design}
     \intone dU U_{i j} U_{m k}^{*}=\frac{\delta_{i m} \delta_{j k}}{n}&,
\end{flalign}
where $\intone dU$ is the integral over the 1-design ensemble of the unitary matrices. 
Second, if the ensemble of $n\times n$ unitary matrices $\{ U \}$ is 2-design, the following 
formula holds:
\begin{flalign}
\label{2-design}
      \inttwo dU U_{i_1 j_1}U_{i_2 j_2} 
          U_{i_1^{\prime}j_1^{\prime}}^{\ast}U_{i_2^{\prime}j_2^{\prime}}^{\ast} 
             = \sum_{k=1}^4 \lambda_{k}^{(n)} 
                   \Delta^k_{i_1 j_1 i_2 j_2 i_1^{\prime}j_1^{\prime} i_2^{\prime}j_2^{\prime}}, 
\end{flalign}
where 
\begin{flalign}
      \lambda_{1}^{(n)} = \lambda_{2}^{(n)} = \frac{1}{2^{2n}-1} &, ~~
      \lambda_{3}^{(n)} = \lambda_{4}^{(n)} = -\frac{1}{(2^{2n}-1)2^n},
\nonumber \\
      \Delta^1_{i_1 j_1 i_2 j_2 i_1^{\prime}j_1^{\prime} i_2^{\prime}j_2^{\prime}} &= \delta_{i_1i_1^{\prime}}\delta_{j_1j_1^{\prime}}\delta_{i_2i_2^{\prime}}\delta_{j_2j_2^{\prime}}, \nonumber\\
		\Delta^2_{i_1 j_1 i_2 j_2 i_1^{\prime}j_1^{\prime} i_2^{\prime}j_2^{\prime}} &= \delta_{i_1i_2^{\prime}}\delta_{j_1j_2^{\prime}}\delta_{i_2i_1^{\prime}}\delta_{j_2j_1^{\prime}}, \nonumber\\
		\Delta^3_{i_1 j_1 i_2 j_2 i_1^{\prime}j_1^{\prime} i_2^{\prime}j_2^{\prime}} &= \delta_{i_1i_1^{\prime}}\delta_{j_1j_2^{\prime}}\delta_{i_2i_2^{\prime}}\delta_{j_2j_1^{\prime}}, \nonumber\\
		\Delta^4_{i_1 j_1 i_2 j_2 i_1^{\prime}j_1^{\prime} i_2^{\prime}j_2^{\prime}} &= \delta_{i_1i_2^{\prime}}\delta_{j_1j_1^{\prime}}\delta_{i_2i_1^{\prime}}\delta_{j_2j_2^{\prime}}, 
\end{flalign}
and $\inttwo dU$ is the integral over the 2-design ensemble of the unitary matrices. 
These formulae are effectively used to derive the theorems shown below.

\subsubsection*{The First Frame Potential}

For the first frame potentials, the following theorem holds.

\begin{th.}
\label{first-moment-1}
If the ensemble of the unitary matrices corresponding to  
$\hea$ and the ensemble of the unitary matrices corresponding to each block of $\alt$ and 
$\ten$ are 2-design, then the following equalities hold: 
\begin{flalign}
\label{first-moment-formula}
     \f{1}{\hea} = \f{1}{\alt} = \f{1}{\ten} = \fhaarn{1}.
\end{flalign}
\end{th.}

The equality, $\f{1}{\hea} = \fhaarn{1}$, can be readily proved  from the assumption of 
the theorem, because, if the ensemble of the unitary matrices corresponding to the circuit 
is 2-design, the ensemble of the states generated by the circuit is a state 2-design (and 
therefore a state 1-design). 
For the other equalities, the proof is given in Appendix. 
Note that, accordingly, the ensembles of the states generated by $\hea$, $\alt$, and $\ten$ 
are all 1-design.

\subsubsection*{The Second Frame Potential}

The second frame potential of the Haar random circuits, $\fhaarn{2}$, can be computed as 
\begin{flalign}
\qquad	
       \fhaarn{2} = \int_0^{1}dF F^2(2^n - 1) (1 - F)^{2^n-2} 
               &= \frac{1}{(2^n + 1)2^{n-1}}.
\end{flalign}
Then, for $\hea$ and  $\ten$, the following theorem holds. 
\begin{th.}
\label{second-moment-1}
If the ensemble of the unitary matrices corresponding to $\hea$ and the ensemble of 
the unitary matrices corresponding to each block of $\ten$ are 2-design, then the following 
equalities hold: 
\begin{flalign}
    \fhea{2} &= \fhaarn{2}, \label{hea-second}\\
    \ften{2} &= 2^{\frac{n}{m}-1} \cdot \frac{2^n + 1}{(2^m + 1)^{\frac{n}{m}}}\fhaarn{2}. 
    \label{ten-second}
\end{flalign}
\end{th.}
The equality (\ref{hea-second}) is readily proved from the assumption of the theorem, because, 
as we mentioned above, if the ensemble of the unitary matrices corresponding to the circuit is 
2-design, the ensemble of the states generated by the circuit is a state 2-design. 
For Eq.\eqref{ten-second}, we give the proof in Appendix. 
From this theorem we find that $\ften{2}$ is always larger than $\fhaarn{2}$; in particular, 
$\ften{2} \simeq 2^{n/m-1} \fhaarn{2}$ for large $n$, meaning that the expressibility 
of TEN is much smaller than that of HEA in the sense of the frame potential.

As for ALT, it is difficult to obtain an explicit formula like the case of HEA and TEN. 
Hence in Theorem~3 below, we provide a formula for computing the values of $\faltcustom{2}{2}$ 
and $\faltcustom{2}{3}$; the computation methods for the other $\ell$s are left for future work. 
Before stating the theorem, below we define a 16-dimensional vector $\atwom$, a $16 \times 16$ 
matrix $B(2, m)$, a 64-dimensional vector $\athreem$, and a $64 \times 64$ matrix $B(3, m)$. 
Given integers $k_a, k_b \in \{1, 2, 3, 4\}$, the $(4(k_a-1) + k_b)$-th component of the vector 
$\atwom$ is defined as
\begin{flalign}
\label{a2m}
	\atwom_{4(k_a-1) + k_b} 
	  &= \inttwo dP dQ \sqrt{\lambda_{k_a}^{(m)}\lambda_{k_b}^{(m)}}\Delta^{(k_a, k_b)}(P, Q), 
\end{flalign}
where $\Delta^{(k_a, k_b)}(P, Q)$ is the function of $m/2 \times m/2$ unitary matrices 
$P$ and $Q$:
\begin{flalign}
\Delta^{(k_a, k_b)}(P, Q) &= \sum_{\substack{u u^{\prime} i i^{\prime}\\
j j^{\prime} l l^{\prime}}}
\sum_{\substack{p p^{\prime} q q^{\prime}}} \Delta^{k_a}_{u 0 u^{\prime} 0 i 0 i^{\prime} 0} \Delta^{k_b}_{p 0 p^{\prime} 0 q 0 q^{\prime} 0} 
		P_{j u} P_{j^{\prime} u^{\prime}} P_{l i}^{\ast} P_{l^{\prime} i^{\prime}}^{\ast} 
		Q_{l p} Q_{l^{\prime} p^{\prime}} Q_{j q}^{\ast} Q_{j^{\prime} q^{\prime}}^{\ast}.
\end{flalign}
Next, given integers $k_a, k_b, k_c \in \{1, 2, 3, 4\}$, the $(16(k_a-1) + 4(k_b-1) + k_c)$-th 
component of the vector $\athreem$ is defined as
\begin{flalign}
\label{a3m}
	\athreem_{16(k_a-1) + 4(k_b-1) + k_c} 
	   &= \inttwo dP dQ \sqrt{\lambda_{k_a}^{(m)}\lambda_{k_b}^{(m)}\lambda_{k_c}^{(m)}}\Delta^{(k_a, k_b, k_c)}(P, Q),
\end{flalign}
where $\Delta^{(k_a, k_b, k_c)}(P, Q)$ is a function of $m/2 \times m/2$ unitary matrices 
$P$ and $Q$:
\begin{flalign}
\label{delta-3}
	\Delta^{(k_a, k_b, k_c)}(P, Q) 
		&= \sum_{\substack{u u^{\prime} i i^{\prime}\\
j j^{\prime} l l^{\prime}}}
\sum_{\substack{r r^{\prime} t t^{\prime}\\
p p^{\prime} q q^{\prime}}}
		\Delta^{k_a}_{u 0 u^{\prime} 0 i 0 i^{\prime} 0} 
		\Delta^{k_b}_{j l j^{\prime} l^{\prime} t r t^{\prime} r^{\prime}} \Delta^{k_c}_{p 0 p^{\prime} 0 q 0 q^{\prime} 0} 
		P_{t u} P_{t^{\prime} u^{\prime}} P_{j i}^{\ast} P_{j^{\prime} i^{\prime}}^{\ast} 
		Q_{l p} Q_{l^{\prime} p^{\prime}} Q_{r q}^{\ast} Q_{r^{\prime} q^{\prime}}^{\ast}.
\end{flalign}
Also, given integers $k_a, k_b, k_c, k_d\in \{1, 2, 3, 4\}$, the $(4(k_a-1) + 4k_b, 4(k_c-1) + k_d)$-th 
component of the matrix $B(m, 2)$ is defined as 
\begin{flalign}
\label{b2m}
	B(2, m)_{4(k_a-1) + k_b, 4(k_c-1) + k_d} &= \inttwo dP dQ \sqrt{\lambda_{k_a}^{(m)}\lambda_{k_b}^{(m)}}\sqrt{\lambda_{k_c}^{(m)}\lambda_{k_d}^{(m)}}\Delta^{(k_a, k_b, k_c, k_d)}(P, Q), 
\end{flalign}
where $\Delta^{(k_a, k_b, k_c, k_d)}(P, Q)$ is a function of $m \times m$ matrices $P$ and $Q$:
\begin{flalign}
\Delta^{(k_a, k_b, k_c, k_d)}(P, Q) &= \sum_{\substack{u_2 u^{\prime}_2 i_2 i^{\prime}_2\\
j_2 j^{\prime}_2 l_2 l^{\prime}_2}}
\sum_{\substack{
p_2 p^{\prime}_2 q_2 q^{\prime}_2}}
\sum_{\substack{u_3 u^{\prime}_3 i_3 i^{\prime}_3\\
j_3 j^{\prime}_3 l_3 l^{\prime}_3}}
\sum_{\substack{
p_3 p^{\prime}_3 q_3 q^{\prime}_3}} \Delta^{k_a}_{u_2 0 u^{\prime}_2 0 i_2 0 i^{\prime}_2 0} \Delta^{k_b}_{p_2 0 p^{\prime}_2 0 q_2 0 q^{\prime}_2 0} \Delta^{k_c}_{u_3 0 u^{\prime}_3 0 i_3 0 i^{\prime}_3 0} \Delta^{k_d}_{p_3 0 p^{\prime}_3 0 q_3 0 q^{\prime}_3 0} \nonumber\\
		&\qquad \times P_{j_2 u_2}^{j_3 u_3} P_{j^{\prime}_2 u^{\prime}_2}^{j^{\prime}_3 u^{\prime}_3} P_{l_2 i_2}^{\ast l_3 i_3} P_{l^{\prime}_2 i^{\prime}_2}^{\ast l^{\prime}_3 i^{\prime}_3} 
		Q_{l_2 p_2}^{l_3 p_3} Q_{l^{\prime}_2 p^{\prime}_2}^{l^{\prime}_3 p^{\prime}_3} Q_{j_2 q_2}^{\ast j_3 q_3} Q_{j_2^{\prime} q^{\prime}_2}^{\ast j_3^{\prime} q^{\prime}_3 }
\end{flalign}
For the matrix component $M^{s, t}_{i, j}$, the upper indices correspond to the first $m/2$ 
qubits and the lower indices correspond to the last $m/2$. 
Given integers $k_a, k_b, k_c, k_d, k_e, k_f \in \{1, 2, 3, 4\}$, 
the $(16(k_a-1) + 4(k_b-1) + k_c, 16(k_d-1) + 4(k_e-1) + k_f)$-th component of the matrix 
$B(m, 3)$ is defined as 
\begin{flalign}
\label{b3m}
	B(3, m)&_{16(k_a-1) + 4(k_b-1) + k_c, 16(k_d-1) + 4(k_e-1) + k_f} 
	 \nonumber \\ 
	 &= \inttwo  dP dQ \sqrt{\lambda_{k_a}^{(m)}\lambda_{k_b}^{(m)}\lambda_{k_c}^{(m)}}\sqrt{\lambda_{k_d}^{(m)}\lambda_{k_e}^{(m)}\lambda_{k_f}^{(m)}}\Delta^{(k_a, k_b, k_c, k_d, k_e, k_f)}(P, Q), 
\end{flalign}
where $\Delta^{(k_a, k_b, k_c, k_d, k_e, k_f)}(P, Q)$ is a function of $m \times m$ matrices 
$P$ and $Q$:
\begin{flalign}
\label{delta-six}
	\Delta^{(k_a, k_b, k_c, k_d, k_e, k_f)}(P, Q) &=\sum_{\substack{u_2 u^{\prime}_2 i_2 i^{\prime}_2\\
j_2 j^{\prime}_2 l_2 l^{\prime}_2}}
\sum_{\substack{r_2 r^{\prime}_2 t_2 t^{\prime}_2\\
p_2 p^{\prime}_2 q_2 q^{\prime}_2}}
\sum_{\substack{u_3 u^{\prime}_3 i_3 i^{\prime}_3\\
j_3 j^{\prime}_3 l_3 l^{\prime}_3}}
\sum_{\substack{r_3 r^{\prime}_3 t_3 t^{\prime}_3\\
p_3 p^{\prime}_3 q_3 q^{\prime}_3}}
\Delta^{k_{11}^a}_{u_20u^{\prime}_20i_20i^{\prime}_20}
\Delta^{k_{31}^a}_{j_2l_2 j^{\prime}_2 l^{\prime}_2 t_2 r_2 t^{\prime}_2 r^{\prime}_2} 
\Delta^{k_{11}^b}_{p_20p^{\prime}_2q_2 0q^{\prime}_20} \nonumber\\
&\qquad \times \Delta^{k_{12}^a}_{u_30u^{\prime}_30i_30i^{\prime}_30}
\Delta^{k_{32}^a}_{j_3l_3 j^{\prime}_3 l^{\prime}_3 t_3 r_3 t^{\prime}_3 r^{\prime}_3} 
\Delta^{k_{12}^b}_{p_30p^{\prime}_3q_3 0q^{\prime}_30}
P_{t_2 u_2}^{t_3 u_3}
P_{t^{\prime}_2 u^{\prime}_2}^{t^{\prime}_3 u^{\prime}_3}
P^{\ast j_3 i_3}_{j_2 i_2}
P^{\ast j^{\prime}_3 i^{\prime}_3}_{j^{\prime}_2 i^{\prime}_2} \nonumber\\
&\qquad \times Q_{l_2 p_2}^{l_3 p_3}
Q_{l^{\prime}_2 p^{\prime}_2}^{l^{\prime}_3 p^{\prime}_3}
Q^{\ast r_3 q_3}_{r_2 q_2}
Q^{\ast r^{\prime}_3 q^{\prime}_3}_{r^{\prime}_2 q^{\prime}_2}.
\end{flalign}
Now we can give the theorem as follows (the proof is given in Appendix). 
\begin{th.}
\label{second-moment-2}
If the ensemble of the unitary matrices corresponding to each block of $\altcustom{2}$ and 
$\altcustom{3}$ are 2-design, then we have
\begin{flalign}
\label{aba2}
      & \faltcustom{2}{2} = \atwom^{\rm T} B(2, m)^{\frac{n}{m} - 1} \atwom, 
\\
\label{aba3}
      & \faltcustom{2}{3} = \athreem^{\rm T} B(3, m)^{\frac{n}{m} - 1} \athreem.
\end{flalign}
%
%
\end{th.}

\fig{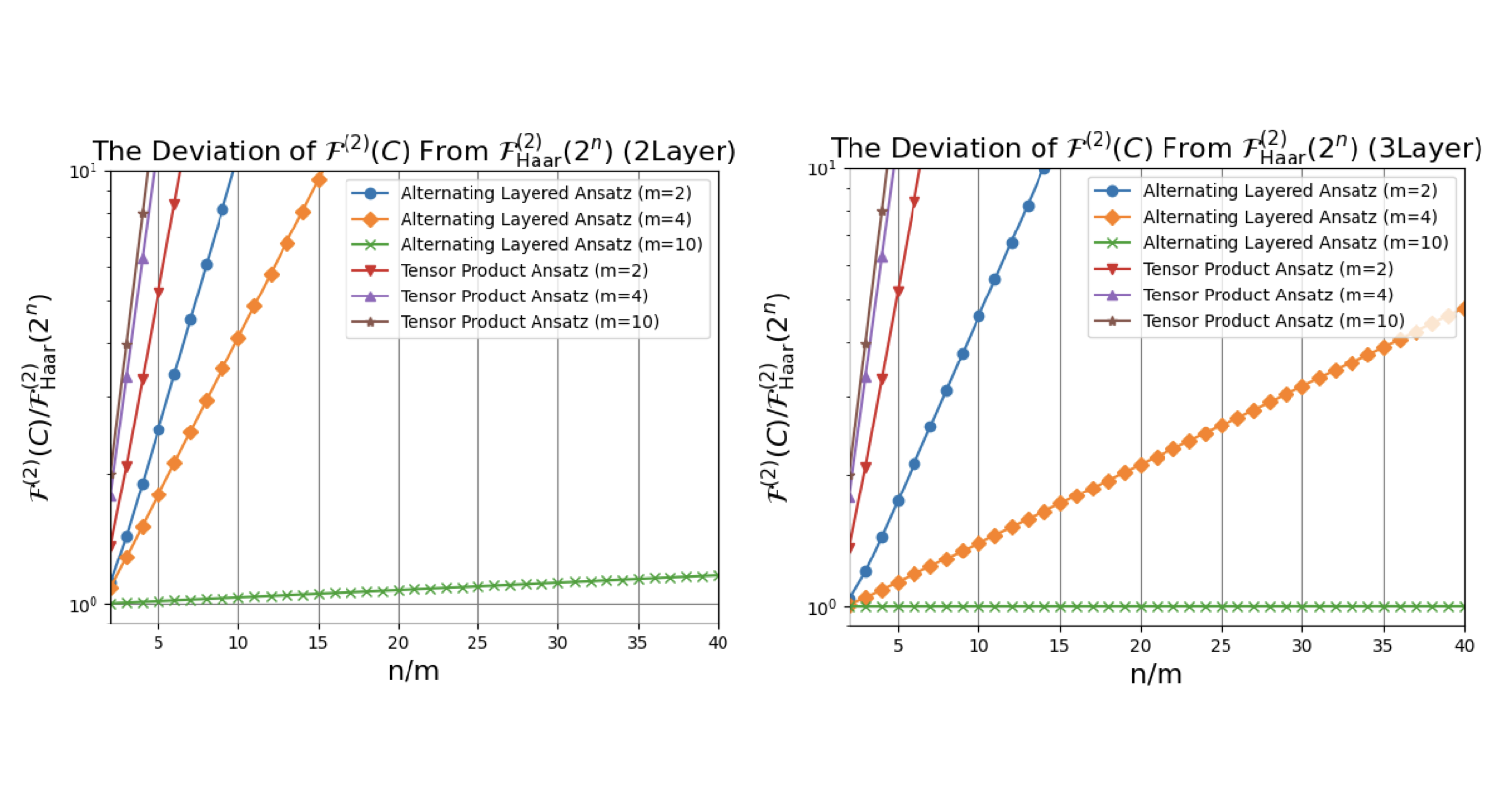}
{The values of $\f{2}{C}/\fhaarn{2}$ for ALT and TEN. 
The left and right figures show the case $\ell = 2$ and $\ell = 3$, respectively. 
The vertical axes are in the logarithmic scale. 
If $\f{2}{C}/\fhaarn{2}$ is close to 1, this means that the expressibility of the ansatz $C$ 
is relatively high.}
{figure-second-moment}{-1}

The vectors $\aellm$ and the matrices $B(\ell, m)$ for $\ell = 2,3$ are obtained by 
directly computing Eqs. (\ref{a2m}), (\ref{a3m}), (\ref{b2m}), and (\ref{b3m}), which then 
lead to $\faltcustom{2}{2}$ and $\faltcustom{2}{3}$. 
Now our interest is in the gap of these quantities from $\fhaarn{t}$, to see the 
expressibility of ALT. 
For this purpose, in Fig.~\ref{figure-second-moment} we show $\faltcustom{2}{2}/\fhaarn{2}$ 
and $\faltcustom{2}{3}/\fhaarn{2}$ as a function of $n/m$, for several values of $(m, n)$. 
For comparison, $\ftencustom{2}{2}/\fhaarn{2}$ and $\ftencustom{2}{3}/\fhaarn{2}$ are 
shown in the figure. 
Recall that, if this measure takes a smaller value, this means that the corresponding ansatz 
has a higher expressibility. 
Here is the list of notable points:
\begin{itemize}
\item
For any pair of $(n, m)$, it is clear that $\faltcustom{2}{\ell}$ is much smaller than 
$\ftencustom{2}{\ell}$ for both $\ell=2,3$. 
This means that, as expected, ALT has a much higher expressibility than TEN. 

\item
For any pair of $(n, m)$,  $\faltcustom{2}{2} > \faltcustom{2}{3}$ hold, i.e., as $\ell$ increases, the expressibility increases.

\item
For any fixed $n/m$, the ALT with bigger $m$ always has a higher expressibility. 
For instance, the ALT with $(n,m)=(50,10)$ has a higher expressibility than the ALT with 
$(n,m)=(20,4)$. 
This is simply because, if the structure of the circuit (the number of division in each layer 
for making the block) is the same, then an ALT with bigger block components has a 
higher expressibility. 

\item
For a fixed $n$, we have ALT with the smaller second order frame potential by taking 
$m$ bigger. 
For instance $n=100$, we have 
${\mathcal F}^{(2)}(C_{\rm ALT}^{2,2,100}) > {\mathcal F}^{(2)}(C_{\rm ALT}^{2,4,100}) 
> {\mathcal F}^{(2)}(C_{\rm ALT}^{2,10,100})$. 
That is, for a limited number of available qubits, the ALT with less blocks has a higher 
expressibility. 

\item
$\faltcustom{2}{\ell} \simeq \fhaarn{2}$ when $m=10$ for all $n/m$ within the figure and for both $\ell=2,3$. 
Hence the ALT composed from the blocks with $m=10$ qubits in each layer has almost the 
same expressibility as the HEA without respect to the total qubits number, $n$. 
In other words, for a given HEA with fixed $n$, we can divide each layer into separated 10-qubits 
blocks to make an ALT, without decreasing the expressibility. 

\end{itemize}
The last point is of particular important in our scenario. 
That is, we are concerned with the condition on the number $m$ such that 
$\faltcustom{2}{\ell} \simeq \fhaarn{2}$ holds. 
The following Theorem~4 and the subsequent Corollary~1, which can be readily derived 
from the theorem, provide a means for evaluating such $m$. 
\begin{th.}
\label{second-moment-alt}
If the ensemble of the unitary matrices corresponding to each block of $\altcustom{2}$ and 
$\altcustom{3}$ are 2-design, then the following inequalities hold: 
\begin{flalign}
	\faltcustom{2}{2} &<  \left(1+\frac{1}{2^{n}}\right)\left(1+\frac{1.2}{2^m}\right)^2\left(1+ 8\left(\left(1 + \frac{20.8}{2^{m/2}}\right)^{\frac{n}{m}-1} -1\right)\right)\fhaarn{2}, \\
	\faltcustom{2}{3} &<  \left(1+\frac{1}{2^{n}}\right)\left(1+\frac{1.2}{2^m}\right)^2\left(1+ 32\left(\left(1 + \frac{83.2}{2^{m/2}}\right)^{\frac{n}{m}-1} -1\right)\right)\fhaarn{2}.
\end{flalign}
\end{th.}
\begin{corollary}
\label{large-n}
If $m=2a\log_2 n$ and $143/(an^{a-1}\log_2 n) < 1$,
\begin{flalign}
	\faltcustom{2}{2} &<  \left(1+\frac{1}{2^{n}}\right)\left(1+\frac{1.2}{n^{2a}}\right)^2\left(1+ \frac{143}{an^{a-1} \log_2 n}\right)\fhaarn{2}. 
\end{flalign}
If $m=2a\log_2 n$ and $2288/(an^{a-1}\log_2 n) < 1$,
\begin{flalign}
	\faltcustom{2}{3} &<  
	\left(1+\frac{1}{2^{n}}\right)\left(1+\frac{1.2}{n^{2a}}\right)^2\left(1+ \frac{2288}{an^{a-1}\log_2 n}\right)\fhaarn{2}.
\end{flalign}
\end{corollary}

Recall from Eq.~\eqref{frame-potential ineq} that $\f{t}{C} \geq \fhaar{t}$ holds for any ansatz $C$. 
Therefore, if $m \geq 4\log n$ and $n$ is enough large, Corollary~1 implies that 
$\faltcustom{2}{2} \sim \fhaarn{2}$ and $\faltcustom{2}{3} \sim \fhaarn{2}$. 
This means that the ensembles of the states generated by  $\altcustom{2}$ 
and $\altcustom{3}$ are almost 2-design. 
Hence in this case, from Theorem~2, the expressibility of ALT is as high as that of HAE. 
It is worth mentioning that, when $m=O(\log_2 n)$, the vanishing gradient problem does 
not happen in ALT as long as the cost function is local and $\ell$ is small \cite{local-cost-function}. 
More precisely, it was shown there that the variance of the gradient of such a cost function 
is larger than the value proportional to $O(1/2^{m\ell})$; thus, by taking $m=O(\log_2 n)$, the 
variance decreases with only $O(1/{\rm poly}(n))$ as a function of $n$, whereas in the 
HEA case the same variance decreases exponentially fast as $n$ becomes large. 
Therefore, the expressibility and the trainability coexists in the shallow ALT with 
$m=O(\log_2 n)$.


\subsection{Expressibility measured by KL divergence}
\label{numerical}

\begin{table}[h]
\small
\begin{center}	
  \begin{tabular}{|p{6mm}|p{12mm}|p{6mm}|p{6mm}|p{18mm}|p{18mm}|p{18mm}|} \hline
    $n$ & {\bf Ansatz} & $\ell$ & $m$ &\begin{tabular}{l}Depth of \\each block \end{tabular} & \begin{tabular}{l}\# of gate\\parameters \end{tabular}&\begin{tabular}{l}Example of\\the circuit \end{tabular}\\ \hline \hline
     \CenterRow{6}{4} & \CenterRow{2}{\bf TEN}& 2 & 2&2&16&-\\ \cline{3-7}
     &  & 3 & 2&2 & 24 &Fig.~\ref{tensor-product-ansatz-ex} \\ \cline{2-7}
      & \CenterRow{2}{\bf ALT} & 2 & 2 & 2 &16&-\\ \cline{3-7}
            &  &3  & 2 & 2 & 24 &Fig.~\ref{alternating-layered-ansatz-ex}\\ \cline{2-7}
       & \CenterRow{1}{\bf HEA} & 4 & - &-&16& Fig.~\ref{hardware-efficient-ansatz-ex}\\ 
      \hline \hline
     \CenterRow{5}{6} & \CenterRow{2}{\bf TEN} &2 & 2 & 2&24 &-\\ \cline{3-7} 
            &  & 3 & 2 & 2 &36&-\\      \cline{2-7}
      & \CenterRow{2}{\bf ALT}& 2 & 2&2&24&-\\ \cline{3-7}
     &  & 3 & 2&2 & 36 &- \\ \cline{2-7}
      & {\bf HEA} & 6 & - & - &36&-\\ \hline \hline
      \CenterRow{10}{8} &  \CenterRow{4}{\bf TEN} & \CenterRow{2}{2} & 2 & 2 & 32 &-\\ \cline{4-7}
            &  &  & 4 & 4 & 64 &-\\ \cline{3-7}
            &  & \CenterRow{2}{3} & 2 & 2 & 48 &-\\ \cline{4-7}
            &  & & 4 & 4 & 96 &-\\ \cline{2-7}
      &  \CenterRow{4}{\bf ALT} & \CenterRow{2}{2} & 2 & 2 & 32 &-\\ \cline{4-7}
            &  &  & 4 & 4 & 64 &-\\ \cline{3-7}
            &  & \CenterRow{2}{3} & 2 & 2 & 48 &-\\ \cline{4-7}
            &  & & 4 & 4 & 96 &-\\ \cline{2-7}
      &  \CenterRow{1}{\bf HEA}  & 8 & - & -&64&-\\ \cline{1-7}
  \end{tabular} 
  \caption{Parameters chosen for computing the KL-divergence. The number of gate parameters are computed by $n \times \ell \times {\rm (Depth\ of\ each\ block)}$ for TEN and ALT, and $n \times \ell$ for HEA.
  }
  \label{kl-setting}
  \end{center}
\end{table}

In Subsection \ref{moment}, we have shown that the first two moments of $\pf{\alt}$ and 
$\pf{\hea}$ are close to those of $\phaar$, as long as $m = O(\log_2 n)$ and the block 
components of the ansatzes are sufficiently random. 
(Recall that, if every block is completely random, then the set of HEA constitutes the Haar 
ensemble.) 
The result implies that both $\pf{\alt}$ and $\pf{\hea}$ are close to $\phaar$ itself, meaning 
that $\pf{\alt} \simeq \pf{\hea} \simeq \phaar$. 
In this subsection, to support this conjecture, we evaluate the values of KL-divergence 
$\expr(C)=D_{\mathrm{KL}}\left(\pf{C} \| P_{\mathrm{Haar}}(F)\right)$ for the case 
$C=\alt$ and $C=\hea$, in addition to $C=\ten$ for comparison with various sets of 
$(\ell, m, n)$. 
Especially, we focus on the relationship between the values of $\f2{C}$ and $\expr(C)$, and 
check if $\f{2}{C} \simeq \f{2}{C^{\prime}}$ would lead to $\expr(C) \simeq \expr(C^{\prime})$ 
for a fixed $n$.

\begin{figure}
  \begin{minipage}[b]{1\linewidth}
    \centering
    \includegraphics[width=330mm,bb=0 0 1558 140]{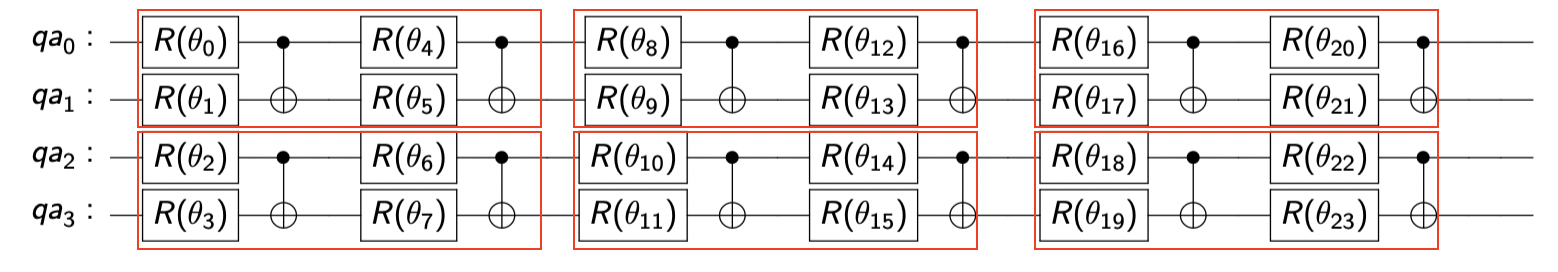}
    \subcaption{The structure of $\tenbase{3}{2}{4}$. The red boxes correspond to blocks.}
    \label{tensor-product-ansatz-ex}
  \end{minipage}
  \begin{minipage}[b]{1\linewidth}
    \centering
    \includegraphics[width=330mm,bb=0 0 1326 140]{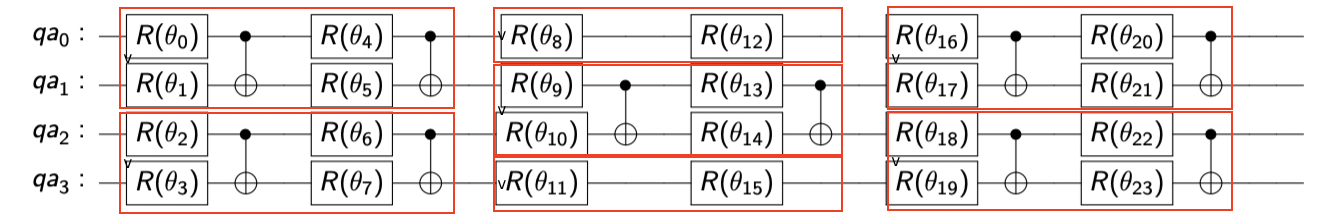}
    \subcaption{The structure of $\altbase{3}{2}{4}$. The red boxes correspond to blocks.}
    \label{alternating-layered-ansatz-ex}
  \end{minipage}
   \begin{minipage}[b]{1\linewidth}
    \centering
    \includegraphics[width=330mm,bb=0 0 1790 200]{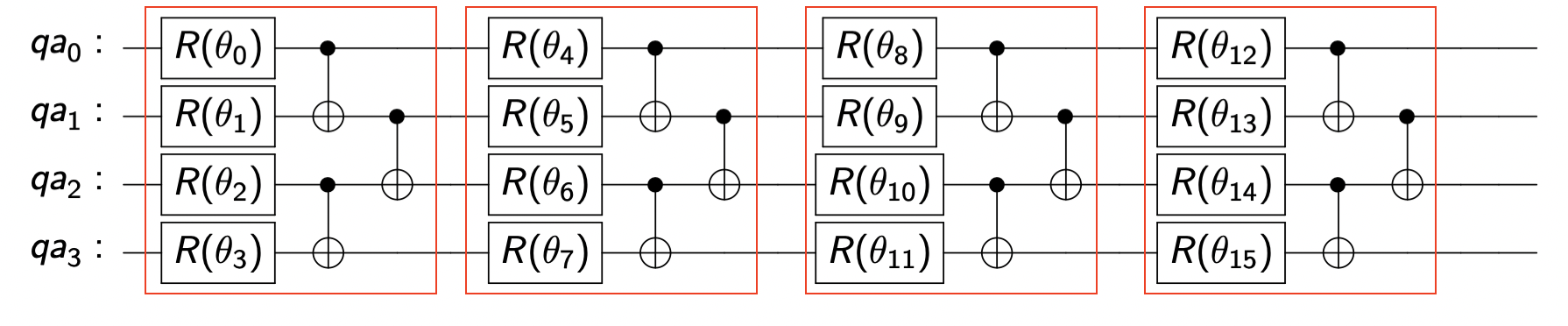}
    \subcaption{The structure of $\heabase{4}{4}$.}
    \label{hardware-efficient-ansatz-ex}
  \end{minipage}
  \caption{The structures of ansatzes in our setting. }
\end{figure}

The parameters taken for calculating the KL-divergence are summarized in Table~\ref{kl-setting}. 
Note that the circuits are chosen to be similar to those used in Section~\ref{moment}; 
for TEN and ALT, the depth of the circuits inside the blocks are all set to $m$ so that the 
ensemble of the unitary matrices corresponding to those circuits become close to
2-design \cite{approximate-2-design,approximate-2-design-2}; for HEA, $\ell$ is set to 
$n$ so that the ensemble of the unitary matrices corresponding to the {\it whole circuits} 
becomes close to 2-design. 
It is expected that $\faltbase{2}{3}{2}{4} \approx \fheabase{2}{4}{4}$ and 
$\faltbase{2}{3}{4}{8} \approx \fheabase{2}{8}{8}$ are realized, because, in 
Fig~\ref{figure-second-moment}, we see that $\faltbase{2}{3}{2}{4}$ and 
$\faltbase{2}{3}{4}{8}$ are almost equals to the Haar values when the ensembles 
of unitary matrices corresponding to each block are 2-design. 
Thus, we here check if $\f{2}{C} \simeq \f{2}{C^{\prime}}$ would mean 
$\expr(C) \simeq \expr(C^{\prime})$ in these parameter sets. 
As an example of the circuit, the whole structure of $\tenbase{3}{2}{4}$, $\altbase{3}{2}{4}$, and 
$\heabase{4}{4}$ in our settings are shown in Figs.~\ref{tensor-product-ansatz-ex}, 
\ref{alternating-layered-ansatz-ex}, and \ref{hardware-efficient-ansatz-ex}, respectively. 
As illustrated in the figures, each layer is composed of parametrized single qubit gates 
and fixed 2-qubit CNOT gates.

In each trial of computing KL-divergence, we generate 200 states. 
When generating a state in each trial, we randomly choose the parameters and the type fo 
single-qubit gate of the circuit. 
That is, for the $i$-th single qubit gate $R_i(\theta_i) = \exp(\sigma_{a_i} \theta_i)$ with 
$a_i = \{x, y, z\}$ and $\theta_i \in [0, 2\pi]$, in each trial all $a_i$ and $\theta_i$ are randomly 
chosen. 
Then 200 fidelity values are computed, which are then used to construct the histogram with 
1000 bins to approximate the probability distribution $\pf{C}$. 
Note that increasing the number of generated states and the number of bins do not affect 
the following conclusions.

In this setting, Fig.~\ref{expr} shows the KL divergences $\expr(\alt)$, $\expr(\hea)$, and 
$\expr(\ten)$. 
As a reference, we also show the values of $\f{2}{C}/\fhaarn{2}$ computed from the 
second moment of the fidelity distributions. 
Each data point and associated error bar is the average and the standard deviation of 
10 trials of computation, respectively. 
Here is the list of points:
\begin{itemize}
\item
For a fixed $n$, $\expr(\ten)$ is always bigger than $\expr(\alt)$ and $\expr(\hea)$. 
\item  
As the number of layers increases, the KL-divergence decreases for fixed $(m, n)$. 
\item
For a fixed $n$, the tendency of the values of $\f{2}{C}/\fhaarn{2}$ is strongly correlated 
with that of KL-divergence. 
\item As expected, $\faltcustom{2}{3} \approx \fheacustom{2}{n}$ is realized when 
$(m, n) = (2, 4)$ and $(4, 8)$. 
\item 
$\expr(\altcustom{3})$ is as small as $\expr(\heabase{n}{n})$ in the parameter sets where 
$\faltcustom{2}{3} \approx \fheacustom{2}{n}$  is realized, i.e., $(m, n) = (2, 4)$ and 
$(4, 8)$. 
This result implies that the state distribution in ALT is also close to that in HEA, in the 
setting where the second frame potential is close to $\fhaarn{2}$. 
\end{itemize}

From some of the above observations, we find the strong correlation between 
$\f{2}{C}/\fhaarn{2}$ and $\expr(C)$; 
that is, as $\f{2}{C}$ becomes close to $1$, then $\expr(C)$ becomes close to $0$. 
Therefore, combining the result obtained in Section~\ref{moment}, we get a clear evidence 
that, as far as $m = O(\log_2 n)$, $\expr(\altcustom{2})\approx 0$ and 
$\expr(\altcustom{3})\approx 0$ hold. 
That is, the high expressibility and trainability in ALT proven in Section~\ref{moment} are 
assured also in terms of KL-divergence.

\fig{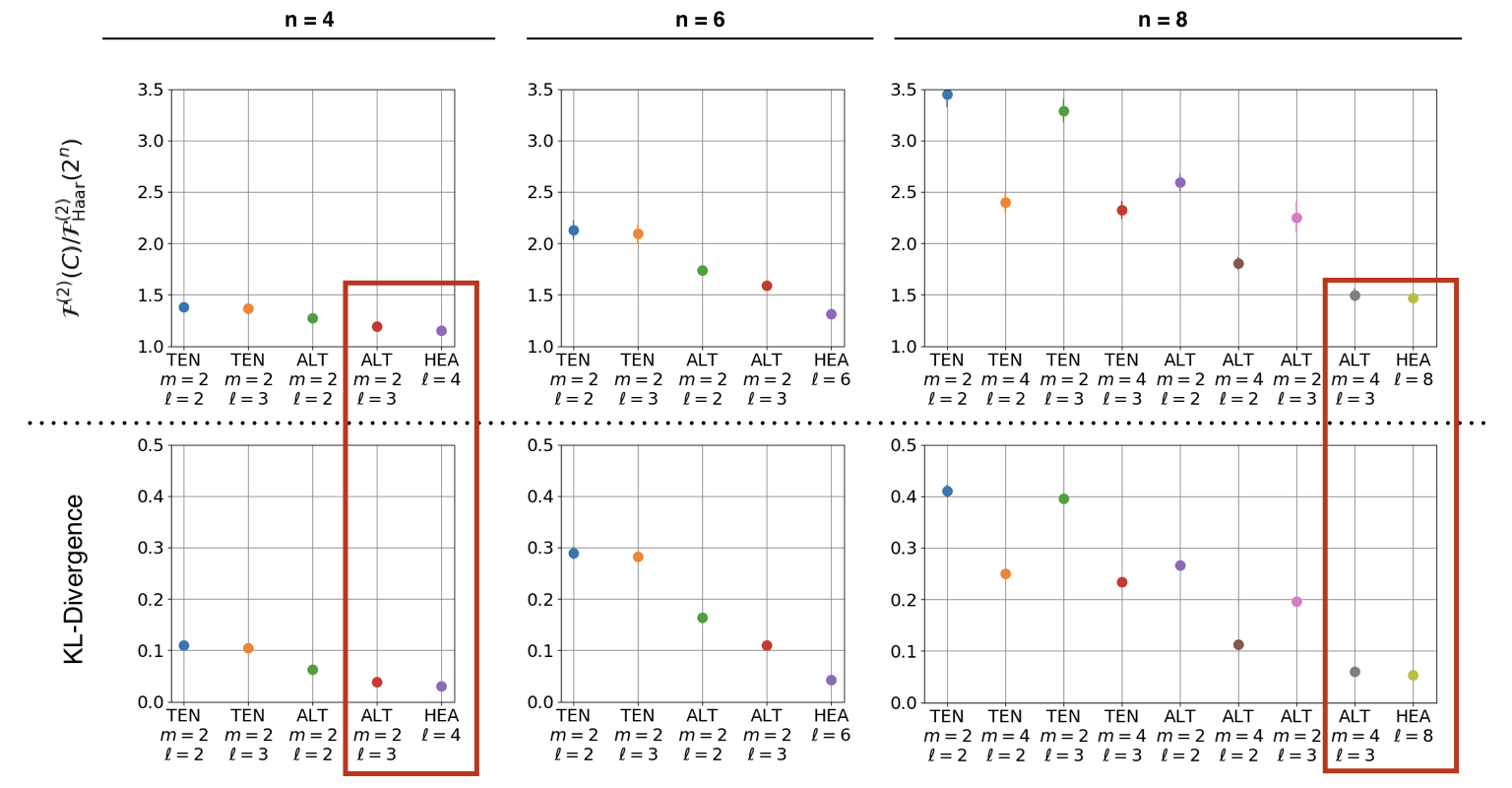}
{$\f{2}{C}/\fhaarn{2}$ (top) and KL-divergence (bottom) for each ansatz. 
The sets of points with which $\faltcustom{2}{3} \approx \fheacustom{2}{3}$ hold 
are enclosed by the red rectangles.}
{expr}{0}


\section{Application to VQE}
\label{vqe-section}

Recall that ALT was originally introduced with the motivation to resolve the vanishing 
gradient problem in VQE, which has been often observed when using HEA; 
then we were concerned with the expressibility of ALT in VQE, meaning that ALT would 
not offer a chance to reach the optimal solution due to the possible loss of expressibility. 
But we now know that this concern has been resolved under some conditions, as 
concluded in the previous section; that is, the expressibility and the trainability coexists 
in the shallow ALT with $m=O(\log_2 n)$. 
This section provides a case study of VQE that implies this desirable fact.

We choose the Hamiltonian of 4-qubits Heisenberg model on a 1-dimensional lattice 
with periodic boundary conditions:
\begin{flalign}
\label{hamiltonian}
     {\cal H} &= \sum_{i=1}^{4}
                 (\sigma_x^i \sigma_x^{i+1} + \sigma_y^{i} \sigma_y^{i+1} + \sigma_z^i \sigma_z^{i+1}), 
\end{flalign}
where $\sigma^i_a$ ($a \in \{x, y, z\}$) is the Pauli matrix that operates on the $i$-th qubit 
and $\sigma^5_a = \sigma^1_a$. 
The goal of VQE problem is to find the minimum eigenvalue of ${\cal H}$, by calculating 
the mean energy 
$\langle {\cal H} \rangle = \langle \psi_{\theta} | {\cal H}| \psi_{\theta}\rangle
= \langle 0 | U_C(\theta)^\dagger {\cal H} U_C(\theta) | 0 \rangle$ via a quantum computer 
and then updating the parameter $\theta\in\Theta$ to decrease $\langle {\cal H} \rangle$ via 
a classical computer, in each iteration. 
As ansatzes, $\tenbase{3}{2}{4}$, $\altbase{3}{2}{4}$, and $\heabase{4}{4}$ are chosen. 
As indicated in Fig.~\ref{expr}, the values of KL divergence corresponding to these ansatzes 
show that $\expr(\tenbase{3}{2}{4}) > \expr(\altbase{3}{2}{4}) \simeq \expr(\heabase{4}{4})$. 
That is, this ALT has the expressibility as high as that of HEA, and further, it is expected to 
enjoy the trainability unlike the HEA.

The simulation results are shown in Fig.~\ref{vqe}. 
In the top three subfigures of Fig.~\ref{vqe}, the blue lines and the associated error bars 
represent the average and the standard deviation of $\langle {\cal H} \rangle$ in total 100 
trials, respectively. 
In each trial, the initial parameters of the ansatz are randomly chosen, and the optimization 
to decrease $\langle {\cal H} \rangle$ in each iteration is performed by using the Adam 
Optimizer with learning rate $0.001$ \cite{adam}. 
The green line shows the theoretical minimum energy (i.e., the ground energy) of ${\cal H}$. 
Also the bottom subfigures of Fig.~\ref{vqe} show three trajectories of $\langle {\cal H} \rangle$ 
(red lines) whose energies at the final iteration step are the smallest three.

The ansatz $\tenbase{3}{2}{4}$, which has the least expressibility in the sense of frame 
potentials and the KL divergence analysis, clearly gives the worst result; its least mean-energy 
is far above from the ground energy. 
This is simply because the state generated via $\tenbase{3}{2}{4}$ cannot represent the 
ground state for any parameter choice. 
The result on $\heabase{4}{4}$ is the second worst, which also does not reach the 
ground energy as in the case of TEN. 
Note that increasing the number of parameters does not change this result; that is, we also 
executed the simulation with $\heabase{4}{6}$ that has the same number of parameters as 
$\altbase{3}{2}{4}$ but did not find a better result than that of $\heabase{4}{4}$. 
On the other hand, $\altbase{3}{2}{4}$ succeeds in finding the ground state; in fact, 5 of the 
total 100 trajectories generated via this ALT reach the ground energy. 
Hence this result implies that, in this example, the expressibility and the trainability coexist 
in ALT, while the latter is lacking in HEA.

This better trainability of ALT than that of HEA could be explained in terms of the ``magnitude" 
of the gradient vector $\nabla_{\theta} \langle {\cal H} \rangle 
= [\partial \langle {\cal H} \rangle/\partial \theta_1, \ldots, \partial \langle {\cal H} \rangle/\partial \theta_P]$, 
where $P$ is the number of parameters. 
Now care should be taken to define an appropriate magnitude, because the focused ansatzes have 
different number of parameters. 
In this work, we regard $\partial \langle {\cal H} \rangle/\partial \theta_p, (p=1, \ldots, P)$ as random 
variables and, based on this view, define the magnitude of $\nabla_{\theta} \langle {\cal H} \rangle$ as 
the mean of the absolute value of those random variables: 
\begin{flalign}
\label{EQUATION-g-definition}
  \| g(\theta)\| = \frac{1}{P}\sum_{p=1}^{P} \left| 
       \frac{\partial \langle {\cal H} \rangle}{\partial \theta_p} \right|. 
\end{flalign}
%
We evaluate the average of the magnitude \eqref{EQUATION-g-definition} over sample trajectories, 
at several values of energy reached through the update of $\theta$. 
For this purpose, let $\theta_C^{(i)}(t)$ denote the vector of parameters of a given ansatz $C$ 
at the $t$-th step (number of iteration) of the $i$-th trajectory, and 
$E_C^{(i)}(t) = \langle 0| U_C(\theta_C^{(i)}(t))^{\dagger}\mathcal{H}U_C(\theta_C^{(i)}(t))|0\rangle$ be 
the energy at $\theta_C^{(i)}(t)$. 
Next, to define the average of $\| g(\theta_C)\|$ over the sample trajectories at the given energy 
value $E$, let $t_E^{(i)}$ be the smallest integer such that the energy of the $i$-th trajectory 
satisfies $E_C^{(i)}(t_E^{(i)}) \leq E$; in other words, $t_E^{(i)}$ represents the number of iteration 
such that the $i$-th trajectory first reaches the value $E$. 
Note that some trajectories may not reach a given value $E$ for all the repetition of $\theta_C$. 
(For example, as seen above, all trajectories of TEN never reached the value $E=-7$.) 
Hence, let ${\cal I}_E$ be the set of index $i$ such that the $i$-th trajectory reaches the value $E$ 
at some point of $t_E^{(i)}$. 
We can now define the average of $\| g(\theta_C)\|$ as
\begin{equation}
\label{EQUATION-definition-of-average}
  \langle\| g(\theta_C)\|\rangle_E 
     = \frac{1}{|{\cal I}_E|}\sum_{i\in{\cal I}_E} \left\|g(\theta_C^{(i)}(t_E^{(i)}))\right\|, 
\end{equation}
where $|{\cal I}_E|$ denotes the size of ${\cal I}_E$. 
Figure~\ref{fig-gradient} shows Eq.~\eqref{EQUATION-definition-of-average} for the specific values 
of $E$ (integers from $-7$ to $0$) for the three ansatzes $\tenbase{3}{2}{4}$, 
$\altbase{3}{2}{4}$, and $\heabase{4}{4}$. 
The standard deviation of the average is indicated by the error bar. 
For instance, $\langle \| g(\theta_C)\| \rangle_E \simeq 5.6$ for the case of orange point (i.e., 
the case of ALT) at $E = -1$ was calculated with $|{\cal I}_E|=100$; actually, all 100 trajectories 
become lower than $E = -1$. 
The figure shows that $\langle\| g(\theta_C)\|\rangle_E$ in ALT is always larger than that in HEA for 
all $E$, and this result is consistent to the theorems given in \cite{local-cost-function}. 
Such a larger gradient vector might enable ALT to circumvent possible flat energy landscapes 
and eventually realize the better trainability than HEA, but further studies are necessary to confirm 
this observation. 
Note that TEN has the largest values of $\langle\| g(\theta_C)\|\rangle_E$ when $E\geq -5$, which 
yet do not lead to the convergence to the global minimum due to the lack of expressibility. 

\fig{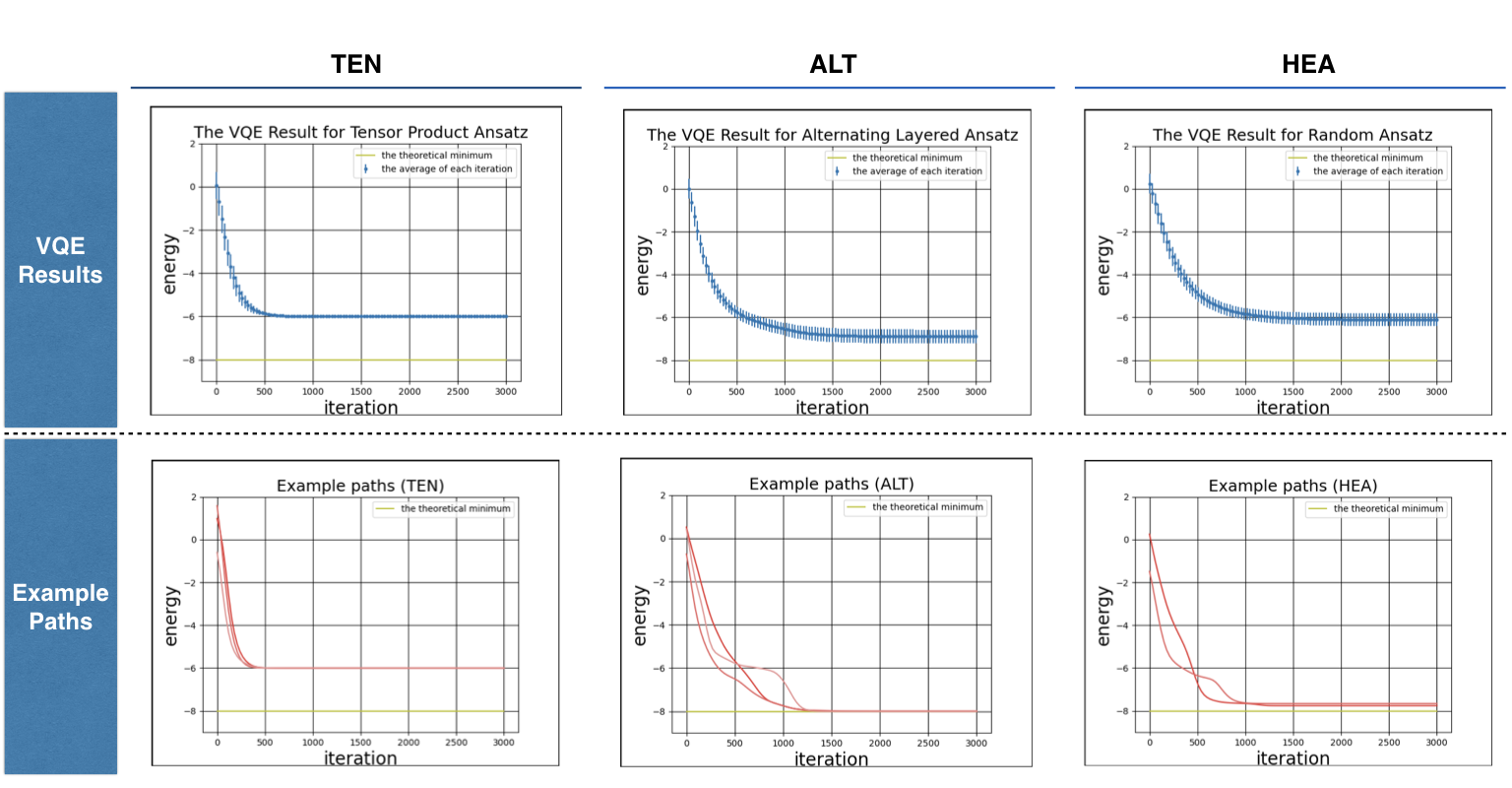}
{Top: Energy versus the iteration step in the VQE problem for the Hamiltonian \eqref{hamiltonian}, 
with the ansatz TEN (left), ALT (center), and HEA (right). 
The blue lines and the associated error bars represent the average and the standard deviation 
of the mean energies in total 100 trials, respectively; 
in each trial, the initial parameters of the ansatz are randomly chosen. 
Optimization to decrease $\langle {\cal H} \rangle$ in each iteration is performed by using 
Adam Optimizer with learning rate $0.001$. 
Bottom: Three of 100 trajectories for each ansatz TEN (left), ALT (center), and HEA (right), 
indicated by red lines. 
The trajectories are chosen such that the energies at the final iteration step are the three 
smallest values. 
}
{vqe}{0}

\fig{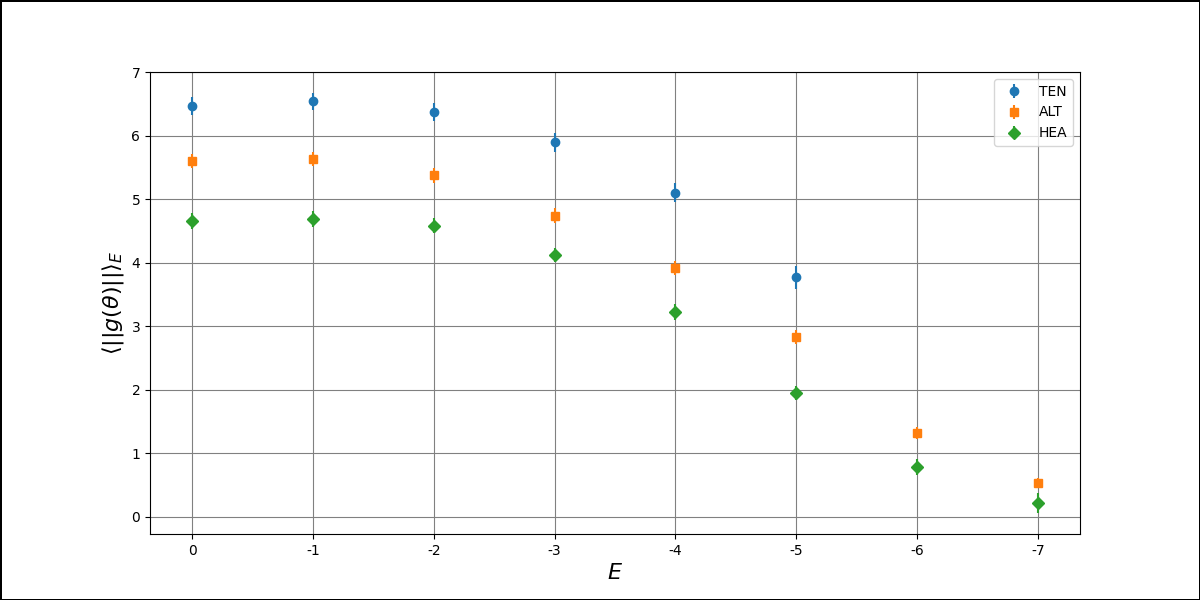}
{The average and the standard deviation of $\|g(\theta)\|$ versus $E$. 
The norm of the gradient is defined as $\| g(\theta)\| =  \frac{1}{P}\sum_{p=1}^{P}\left| \frac{\partial\langle{\mathcal{H}\rangle}}{\partial\theta_p} \right|$.
}
{fig-gradient}
{0}


\section{Conclusion}
\label{conclusion}

This paper has examined the expressibility power of the shallow ALT, which was proposed 
before as a solution to the vanishing gradient problem found in various types of hybrid 
quantum-classical algorithms. 
Our conclusion is that, in addition to such a well trainability, shallow ALTs have the 
expressibility; that is, in the measure of the frame potential and the KL-divergence, those 
shallow ALTs have almost the same expressibility as HEAs, which are often used in hybrid 
algorithms but suffer from the vanishing gradient problem. 
In particular, we have proven that such expressibility holds if the number of entangled 
qubits in each block is of the order of the logarithm of the number of all resource qubits, 
which is consistent to the previous result discussing the trainability of ALT. 
We also provided a case study of VQE problem implying that the ALT enjoys both the 
expressibility and the trainability.

Although our results are limited to the case $\ell = 2, 3$, we have numerically observed 
that the ALT acquires even higher expressibility when making $\ell$ bigger. 
Therefore, we conjecture that the above conclusion still holds for ALT with $\ell \geq 4$. 
The rigorous proof is left for future work.
\\
\mbox{}
{\bf Acknowledgement: } 
This work is supported by the MEXT Quantum Leap Flagship Program 
Grant Number JPMXS0118067285.


\appendix

\section{Proof of Theorems}
\label{proof-of-theorem}

Notation: For the unitary matrix $U_a$ corresponding to the entire circuit, we denote the 
unitary matrix corresponding to the $i$-th layer of the circuit to be $\ualayer{i}$ and 
the unitary matrix corresponding to the $j$-th block in the $i$-th layer as $\uablock{i}{j}$.

\subsection{Proof of Theorem \ref{first-moment-1}}

First, because the probability distribution of $F=|\langle\psi|\psi^{\prime}\rangle|^2$ with 
$n$-qubits states $|\psi\rangle$ and $|\psi^{\prime}\rangle$ taken from the Haar measure 
is given by $P_{\mathrm {Haar}} (F) = (2^n-1) (1-F)^{2^n-2}$, the value of $\fhaarn{1}$ 
is straightforwardly computed as
\begin{flalign}
     \fhaarn{1} 
      = \int_{\rm Haar}\int_{\rm Haar}|\langle\psi|\psi^{\prime}\rangle|^{2}d\psi d\psi^{\prime}
      = \int_0^{1}dF F(2^n - 1) (1 - F)^{2^n-2} = \frac{1}{2^n}.
\end{flalign}
Next, we provide the proof of $\f{1}{\alt} = \fhaarn{1}$. 
Given two final states $ | \phi \rangle = U_{a} |0\rangle$ and 
$ | \psi \rangle = U_{b} | 0 \rangle $ generated by $\alt$, we have 
\begin{flalign}
\label{first-frame-potential}
\qquad	
     \f{1}{\alt} & = \intone d{U_{a}} d{U_{b}} 
         \langle 0 |U_{a}^{\dagger} U_{b} |0\rangle \langle 0|U_{b}^{\dagger} U_{a} |0\rangle 
\nonumber\\
     &= \intone \left(\prod_{i=1}^\ell d\ualayer{i}\right)\left(\prod_{i=1}^\ell d\ublayer{i}\right) \nonumber \\
     &\qquad \times \langle 0 |\ualayer{1}^\dagger \ualayer{2}^\dagger \cdots \ualayer{\ell}^\dagger 
             \ublayer{\ell} \cdots \ublayer{2} \ublayer{1}|0\rangle  
\nonumber\\
     &\qquad \times \langle 0 |\ublayer{1}^\dagger \ublayer{2}^\dagger \cdots \ublayer{\ell}^\dagger 
             \ualayer{\ell} \cdots \ualayer{2} \ualayer{1}|0\rangle 
\nonumber\\
     &= \intone \left(\prod_{i=1}^\ell \left(\prod_{j=1}^{k(i)} d\uablock{i}{j} \right)\right)
                \left(\prod_{i^{\prime}=1}^\ell 
                     \left(\prod_{j^{\prime}=1}^{k(i^{\prime})} d\ubblock{i^{\prime}}{j^{\prime}} \right)\right) 
\nonumber\\ 
&\qquad \times \langle 0 |\ualayer{1}^\dagger \ualayer{2}^\dagger \cdots \ualayer{\ell}^\dagger \ublayer{\ell} \cdots \ublayer{2} \ublayer{1}|0\rangle \nonumber\\
&\qquad \times \langle 0 |\ublayer{1}^\dagger \ublayer{2}^\dagger \cdots \ublayer{\ell}^{\dagger} \ualayer{\ell} \cdots \ualayer{2} \ualayer{1}|0\rangle, 
\end{flalign}
where $k(i)$ is the number of blocks in the $i$-th layer and each $\int dU$ is the average 
over the ensemble of the unitary matrix $U$. 
Because the distribution of each $\uablock{i}{j}$ is 2-design (and is therefore 1-design), 
we can apply the formula (\ref{one-design}) to the integrals with respect to $\uablock{i}{j}$. 
Actually, by integrating $\prod_{j=1}^{k(\ell)} d \ualayer{\ell, j}$ for all $a$ in the last line 
of (\ref{first-frame-potential}), we have 
\begin{flalign}
\label{first-frame-potential-2}
\qquad	
\f{1}{\alt} 
&= \intone \left(\prod_{i=1}^\ell \left(\prod_{j=1}^{k(i)} d\uablock{i}{j} \right)\right)\left(\prod_{i^{\prime}=1}^\ell \left(\prod_{j^{\prime}=1}^{k(i^{\prime})} d\ubblock{i^{\prime}}{j^{\prime}} \right)\right) \nonumber\\ 
&\qquad \times \langle 0 |\ualayer{1}^\dagger \ualayer{2}^\dagger \cdots \ualayer{\ell - 1}^\dagger \ualayer{\ell-1} \cdots \ualayer{2} \ualayer{1}|0\rangle \nonumber\\
&\qquad \times \langle 0 |\ublayer{1}^\dagger \ublayer{2}^\dagger \cdots \ublayer{\ell}^\dagger \ublayer{\ell} \cdots \ublayer{2} \ublayer{1}|0\rangle \nonumber\\
& = \left(\frac{1}{2^m}\right)^{\ell}\intone \left(\prod_{i=1}^{L-1} \left(\prod_{\alpha=1}^{k(i)} dU_a^{i\alpha} \right)\right)\left(\prod_{j=1}^L \left(\prod_{\beta=1}^{k(i^{\prime})} dU_b^{j\beta} \right)\right) \times 1 \nonumber\\
& = \frac{1}{2^n} = \fhaarn{1}.
\end{flalign}
The other equality in Eq.~(\ref{first-moment-formula}) can be proved in the same manner.


\subsection{Proof of Theorem \ref{second-moment-1}}

Similar to the first frame potential, the value of $\fhaarn{1}$ is straightforwardly computed as
\begin{flalign}
\fhaarn{2} = \int_0^{1}dF F^2(2^n - 1) (1 - F)^{2^n-2} = \frac{1}{2^{n-1}(2^n + 1)}.
\end{flalign}
Next, we compute $\ften{2}$. Given two final states $ | \phi \rangle = U_a|0\rangle$ and 
$ | \psi \rangle = U_b |0\rangle$, it is computed as 
\begin{flalign}
	\ften{2} &=  \inttwo \left(\prod_{i=1}^{\ell} d{\uablock{i}{1}} d{\ubblock{i}{1}}\right) |\langle 0 |\uablock{1}{1}^{\dagger} \uablock{2}{1}^{\dagger} \cdots\uablock{\ell}{1}^{\dagger} \ubblock{\ell}{1}\cdots\ubblock{2}{1}\ubblock{1}{1} |0\rangle|^4 
	\nonumber \\
	&\times \inttwo \left(\prod_{i=1}^{\ell} d{\uablock{i}{2}} d{\ubblock{i}{2}}\right)  |\langle 0 |\uablock{1}{2}^{\dagger} \uablock{2}{2}^{\dagger} \cdots\uablock{\ell}{2}^{\dagger} \ubblock{\ell}{2}\cdots\ubblock{2}{2}\ubblock{1}{2} |0\rangle|^4   \nonumber \\
	&\times \cdots \times
	 \inttwo \left(\prod_{i=1}^{\ell} d{\uablock{i}{\frac{n}{m}}} d{\ubblock{i}{\frac{n}{m}}}\right) 
	 \nonumber \\
	 & \qquad\qquad \times |\langle 0 |\uablock{\ell}{1}^{\dagger} \uablock{\ell}{2}^{\dagger} \cdots\uablock{\ell}{\ell}^{\dagger} \ubblock{\ell}{\ell}\cdots\ubblock{\ell}{2}\ubblock{\ell}{1} |0\rangle|^4 
	 	\nonumber \\
	&= \left(\frac{1}{(2^m+1)2^{m-1}}\right)^{\frac{n}{m}} = 2^{\frac{n}{m}-1} \cdot \frac{2^n + 1}{(2^m + 1)^{\frac{n}{m}}}\fhaarn{2}.
\end{flalign}

\subsection{Proof of Theorem \ref{second-moment-2}}

Here we only show the computation of $\faltcustom{2}{3}$. 
The computation of $\faltcustom{2}{2}$ can be done in a similar manner. 
The second frame potential can be expressed as follows:
\begin{flalign*}
    \faltcustom{2}{3} &= \int d\ualayer{1}d\ualayer{2} d\ualayer{3} d\ublayer{1}
      d\ublayer{2} d\ublayer{3} 
      ~ |\langle 0 |\ualayer{1}^{\dagger} \ualayer{2}^{\dagger} \ualayer{3}^{\dagger} 
               \ublayer{3} \ublayer{2}\ublayer{1}|0\rangle|^4 \\
     &= \inttwo \left(\prod_{i=1}^{3}\prod_{j=1}^{k(i)} d\uablock{i}{j}\right)
                  \left(\prod_{i^{\prime}=1}^{3}\prod_{j^{\prime}=1}^{k(i)} 
                        d\ubblock{i^{\prime}}{j^{\prime}}\right) \\
     & \qquad \times  |\langle 0 |\uablock{1}{1}^{\dagger}\uablock{1}{2}^{\dagger} 
            \cdots \uablock{1}{n/m}^{\dagger}\uablock{2}{1}^{\dagger}\uablock{2}{2}^{\dagger}
            \cdots\uablock{2}{n/m+1}^{\dagger} \\
     &\qquad \qquad \times \uablock{3}{1}^{\dagger}\uablock{3}{2}^{\dagger}
             \cdots \uablock{3}{n/m}^{\dagger} \ubblock{3}{n/m}\cdots\ubblock{3}{2}\ubblock{3}{1}
	    \\
      &\qquad\qquad\qquad \times \ubblock{2}{n/m+1}
             \cdots\ubblock{2}{2}\ubblock{2}{1}\ubblock{1}{n/m}
             \cdots\ubblock{1}{2}\ubblock{1}{1} |0\rangle|^4 \nonumber.
\end{flalign*}
Recall that $k(i)$ denotes the number of blocks in $i$-th layer; 
$k(i)=n/m$ for $i=1,3$ and $k(i)=n/m+1$ for $i=2$. 
We can get the final formula in the theorem by integrating only the unitary matrices 
in the first layer and the third layer. 
Executing integrals $\inttwo \prod_{j=1}^{n/m} d\uablock{\ell}{j}$ and 
$\inttwo \prod_{j^{\prime}=1}^{n/m} d\ubblock{\ell}{j^{\prime}}$ for $\ell=1,3$, we have 
\begin{flalign}
\label{integrate-blocks}
     \faltcustom{2}{3} & = \inttwo \left(\prod_{j=1}^{n/m+1} d\uablock{2}{j}\right)
             \left(\prod_{j^{\prime}=1}^{n/m+1} d\ubblock{2}{j^{\prime}}\right)
\nonumber \\
      & \sumak{11}\sumak{12}\cdots\sumak{1\frac{n}{m}}\sumak{31}\sumak{32}\cdots\sumak{3\frac{n}{m}}\sumbk{11}\sumbk{12}\cdots\sumbk{1\frac{n}{m}} \nonumber\\
	    &\lambdaa{11}\lambdaa{12}\cdots
	    \lambdaa{1\frac{n}{m}-1}\lambdaa{1\frac{n}{m}}
	    \lambdaa{31}\lambdaa{32}\cdots
	    \lambdaa{3\frac{n}{m}-1}
	    \lambdaa{3\frac{n}{m}}
	    \lambdab{11}
	    \lambdab{12}\cdots
	    \lambdab{1\frac{n}{m}-1}
	    \lambdab{1\frac{n}{m}}
	    \nonumber\\
		&\times \delthree{1} \times 
		 \delsix{1}{2} \nonumber \\
		 &\times \delsix{2}{3}\times
		 \cdots 
\nonumber \\ 
		  & \times\delsix{\frac{n}{m}-1}{\frac{n}{m}}\nonumber \\
		  & \times \delthreeplus{\frac{n}{m}} \allowdisplaybreaks\nonumber \\
		 &=  \sumak{11}\sumak{12}\cdots\sumak{1\frac{n}{m}}\sumak{31}\sumak{32}\cdots\sumak{3,\frac{n}{m}}\sumbk{11}\sumbk{12}\cdots\sumbk{1\frac{n}{m}}\nonumber \\
		  &\inttwo d\uablock{2}{1} d\ubblock{2}{1}\sqrt{\lambdaa{11}\lambdaa{31}\lambdab{11}}\delthree{1} \nonumber \\ 
		  &\times \inttwo d\uablock{2}{2}d\ubblock{2}{b} \sqrt{\lambdaa{11}\lambdaa{31}\lambdab{11}} 
		  \sqrt{\lambdaa{12}\lambdaa{32}\lambdab{12}} \delsix{1}{2} \nonumber \\ 
		  &\times \inttwo d\uablock{2}{3}d\ubblock{2}{3}\sqrt{\lambdaa{12}\lambdaa{32}\lambdab{12}} 
		  \sqrt{\lambdaa{13}\lambdaa{33}\lambdab{13}}\delsix{2}{3} \nonumber \\
		  &\times \cdots  \times  \inttwo d\uablock{2}{\frac{n}{m}}d\ubblock{2}{\frac{n}{m}}\sqrt{\lambdaa{1\frac{n}{m}-1}\lambdaa{3\frac{n}{m}-1}\lambdab{1\frac{n}{m}-1}} 
		  \sqrt{\lambdaa{1\frac{n}{m}}\lambdaa{3\frac{n}{m}}\lambdab{1\frac{n}{m}}} \nonumber\\
		  &\qquad\qquad \times \delsix{\frac{n}{m}-1}{\frac{n}{m}}
		  \nonumber\\
		  &\times\inttwo d\uablock{2}{\frac{n}{m}+1}d\ubblock{2}{\frac{n}{m}+1} \sqrt{\lambdaa{1\frac{n}{m}}\lambdaa{3\frac{n}{m}}\lambdab{1\frac{n}{m}}}
		  \nonumber\\
		  &\qquad\qquad\times \delthreeplus{\frac{n}{m}}  \nonumber \\
		  &= \athreem^{\rm T} B(3, m)^{\frac{n}{m}-1} \athreem,
\end{flalign}
where we use definitions: (\ref{a3m}), (\ref{delta-3}), (\ref{b3m}), and (\ref{delta-six}).
For the purpose of exemplifying the computation in the first equality of (\ref{integrate-blocks}), 
we show the computation when $n/m=2$ in the following. 
When $n/m=2$, the second frame potential can be computed as follows:
\begin{flalign}
\label{example-second-frame}
\faltbase{2}{3}{n}{m} &= \inttwo dU_a(1,1) dU_a(1,2) dU_a(2,1) dU_a(2,2)dU_a(2,3)dU_a(3,1)dU_a(3,2)\nonumber\\
&\qquad\qquad\qquad dU_b(1,1) dU_b(1,2) dU_b(2,1)dU_b(2,2)dU_b(2,3)dU_b(3,1)dU_b(3,2)\nonumber\\
&\qquad\qquad\qquad|\langle0|U^{\dagger}_a(1,1)U^{\dagger}_a(1,2)U^{\dagger}_a(2,1)U^{\dagger}_a(2,2)U^{\dagger}_a(2,3)U^{\dagger}_a(3,1)U^{\dagger}_a(3,2) \nonumber \\
&\qquad\qquad\qquad\qquad U_b(3,1)U_b(3,2)U_b(2,1)U_b(2,2)U_b(2,3)U_b(1,1)U_b(1,2)\allowdisplaybreaks
|0\rangle|^4 \nonumber\\
&=\inttwo dU_a(1,1) dU_a(1,2) dU_a(2,1) dU_a(2,2)dU_a(2,3)dU_a(3,1)dU_a(3,2)\nonumber\\
&\qquad\qquad\qquad dU_b(1,1) dU_b(1,2) dU_b(2,1)dU_b(2,2)dU_b(2,3)dU_b(3,1)dU_b(3,2)\nonumber\\
&\qquad\qquad\sum_{\bf i,j,k,l,p} \left(U_a^{\ast}(1, 1)_{i_1 0}^{i_2 0}U_a^{\ast}(1, 2)_{i_3 0}^{i_4 0}U_a^{\ast}(2, 1)_{j_1 i_1}U_a^{\ast}(2, 2)_{j_2 i_2}^{j_3 i_3}U_a^{\ast}(2, 3)_{j_4 i_4}
U_a^{\ast}(3, 1)_{k_1 j_1}^{k_2 j_2} U_a^{\ast}(3, 2)_{k_3 j_3}^{k_4 j_4}\right.\nonumber\\
&\qquad\qquad\qquad\left. U_b(3, 1)_{k_1 l_1}^{k_2 l_2}U_b(3, 2)_{k_3 l_3}^{k_4 l_4}U_b(2, 1)_{l_1 p_1}U_b(2, 2)_{l_2 p_2}^{l_3 p_3}U_b(2, 3)_{l_4 p_4}U_b(1, 1)_{p_1 0}^{p_2 0}U_b(1, 2)_{p_3 0}^{p_4 0}\right)\times \nonumber\\
&\qquad\qquad\sum_{\bf q,r,s,t,u} \left(U_b^{\ast}(1, 1)_{q_1 0}^{q_2 0}U_b^{\ast}(1, 2)_{q_3 0}^{q_4 0}U_b^{\ast}(2, 1)_{r_1 q_1}U_b^{\ast}(2, 2)_{r_2 q_2}^{r_3 q_3}U_b^{\ast}(2, 3)_{r_4 q_4}
U_b^{\ast}(3, 1)_{s_1 r_1}^{s_2 r_2} U_b^{\ast}(3, 2)_{s_3 r_3}^{s_4 r_4}\right.\nonumber\\
&\qquad\qquad\qquad\left. U_a(3, 1)_{s_1 t_1}^{s_2 t_2}U_a(3, 2)_{s_3 t_3}^{s_4 t_4}U_a(2, 1)_{t_1 u_1}U_a(2, 2)_{t_2 u_2}^{t_3 u_3}U_a(2, 3)_{t_4 u_4}
U_a(1, 1)_{u_1 0}^{u_2 0}U_a(1, 2)_{u_3 0}^{u_4 0}\right)\times\nonumber\\
&\qquad\qquad\sum_{\bf i^{\prime},j^{\prime},k^{\prime},l^{\prime},p^{\prime}} \left(U_a^{\ast}(1, 1)_{i^{\prime}_1 0}^{i^{\prime}_2 0}U_a^{\ast}(1, 2)_{i^{\prime}_3 0}^{i^{\prime}_4 0}U_a^{\ast}(2, 1)_{j^{\prime}_1 i^{\prime}_1}U_a^{\ast}(2, 2)_{j^{\prime}_2 i^{\prime}_2}^{j^{\prime}_3 i^{\prime}_3}U_a^{\ast}(2, 3)_{j^{\prime}_4 i^{\prime}_4}
U_a^{\ast}(3, 1)_{k^{\prime}_1 j^{\prime}_1}^{k^{\prime}_2 j^{\prime}_2} U_a^{\ast}(3, 2)_{k^{\prime}_3 j^{\prime}_3}^{k^{\prime}_4 j^{\prime}_4}\right. \nonumber\\
&\qquad\qquad\qquad\left. U_b(3, 1)_{k^{\prime}_1 l^{\prime}_1}^{k^{\prime}_2 l^{\prime}_2}U_b(3, 2)_{k^{\prime}_3 l^{\prime}_3}^{k^{\prime}_4 l^{\prime}_4}U_b(2, 1)_{l^{\prime}_1 p^{\prime}_1}U_b(2, 2)_{l^{\prime}_2 p^{\prime}_2}^{l^{\prime}_3 p^{\prime}_3}
U_b(2, 3)_{l^{\prime}_4 p^{\prime}_4}
U_b(1, 1)_{p^{\prime}_1 0}^{p^{\prime}_2 0}U_b(1, 2)_{p^{\prime}_3 0}^{p^{\prime}_4 0}\right)\times \nonumber\\ 
&\qquad\qquad\sum_{\bf q^{\prime},r^{\prime},s^{\prime},t^{\prime},u^{\prime}} \left(U_b^{\ast}(1, 1)_{q^{\prime}_1 0}^{q^{\prime}_2 0}U_b^{\ast}(1, 2)_{q^{\prime}_3 0}^{q^{\prime}_4 0}U_b^{\ast}(2, 1)_{r^{\prime}_1 q^{\prime}_1}U_b^{\ast}(2, 2)_{r^{\prime}_2 q^{\prime}_2}^{r^{\prime}_3 q^{\prime}_3}U_b^{\ast}(2, 3)_{r^{\prime}_4 q^{\prime}_4}
U_b^{\ast}(3, 1)_{s^{\prime}_1 r^{\prime}_1}^{s^{\prime}_2 r^{\prime}_2} U_b^{\ast}(3, 2)_{s^{\prime}_3 r^{\prime}_3}^{s^{\prime}_4 r^{\prime}_4}\right.\nonumber\\
&\qquad\qquad\qquad\left. U_a(3, 1)_{s^{\prime}_1 t^{\prime}_1}^{s^{\prime}_2 t^{\prime}_2}
U_a(3, 2)_{s^{\prime}_3 t^{\prime}_3}^{s^{\prime}_4 t^{\prime}_4}
U_a(2, 1)_{t^{\prime}_1 u^{\prime}_1}
U_a(2, 2)_{t^{\prime}_2 u^{\prime}_2}^{t^{\prime}_3 u^{\prime}_3}
U_a(2, 3)_{t^{\prime}_4 u^{\prime}_4}
U_a(1, 1)_{u^{\prime}_1 0}^{u^{\prime}_2 0}U_a(1, 2)_{u^{\prime}_3 0}^{u^{\prime}_4 0}\right)
\end{flalign}
where the bold symbols in the bottom the summation denote the multiple indices, e.g., ${\bf i}=i_1,i_2,i_3,i_4$.
For the integrals $U_a(1,1), U_b(1,1), U_a(1, 2), U_b(1, 2)$, 
\begin{flalign}
\label{int1}
\inttwo dU_a(1, 1)
U_a(1, 1)_{u_1 0}^{u_2 0}
U_a(1, 1)_{u^{\prime}_1 0}^{u^{\prime}_2 0}
U_a^{\ast}(1, 1)_{i_1 0}^{i_2 0}
U_a^{\ast}(1, 1)_{i^{\prime}_1 0}^{i^{\prime}_2 0}
&= \sum_{k_{11}^a=1}^4 \lambdaa{11}\Delta^{k_{11}^a}_{u_10u^{\prime}_10i_1 0i^{\prime}_10}
\Delta^{k_{11}^a}_{u_20u^{\prime}_20i_2 0i^{\prime}_20}, \\
\label{int2}
\inttwo
dU_b(1, 1)
U_b(1, 1)_{p_1 0}^{p_2 0}
U_b(1, 1)_{p^{\prime}_1 0}^{p^{\prime}_2 0}
U_b^{\ast}(1, 1)_{q_1 0}^{q_2 0}
U_b^{\ast}(1, 1)_{q^{\prime}_1 0}^{q^{\prime}_2 0}
&= \sum_{k_{11}^b=1}^4 \lambdab{11}\Delta^{k_{11}^b}_{p_10p^{\prime}_10q_1 0q^{\prime}_10}
\Delta^{k_{11}^b}_{p_20p^{\prime}_20q_2 0q^{\prime}_20}, \allowdisplaybreaks\\
\label{int3}
\inttwo 
dU_a(1, 2)
U_a(1, 2)_{u_3 0}^{u_4 0}
U_a(1, 2)_{u^{\prime}_3 0}^{u^{\prime}_4 0}
U_a^{\ast}(1, 2)_{i_3 0}^{i_4 0}
U_a^{\ast}(1, 2)_{i^{\prime}_3 0}^{i^{\prime}_4 0}
&= \sum_{k_{12}^a=1}^4 \lambdaa{12}\Delta^{k_{12}^a}_{u_30u^{\prime}_30i_3 0i^{\prime}_30}
\Delta^{k_{12}^a}_{u_40u^{\prime}_40i_4 0i^{\prime}_40}, \\
\label{int4}
\inttwo
dU_b(1, 2)
U_b(1, 2)_{p_3 0}^{p_4 0}
U_b(1, 2)_{p^{\prime}_3 0}^{p^{\prime}_4 0}
U_b^{\ast}(1, 2)_{q_3 0}^{q_4 0}
U_b^{\ast}(1, 2)_{q^{\prime}_3 0}^{q^{\prime}_4 0}
&= \sum_{k_{12}^b=1}^4 \lambdab{12}\Delta^{k_{12}^b}_{p_30p^{\prime}_30q_3 0q^{\prime}_30}
\Delta^{k_{12}^b}_{p_40p^{\prime}_40q_4 0q^{\prime}_40}
\end{flalign}
hold. For the integrals $U_a(3, 1), U_b(3, 1), U_a(3, 2), U_b(3, 2)$, 
\begin{flalign}
\label{int5}
&\inttwo dU_a(3,1)dU_b(3,1) \nonumber\\
&\qquad\sum_{\substack{k_1,k_2\\s_1,s_2\\}}
\sum_{\substack{k^{\prime}_1,k^{\prime}_2\\s^{\prime}_1,s^{\prime}_2\\}} 
U^{\ast}_a(3, 1)^{k_2j_2}_{k_1 j_1} U_b(3, 1)_{k_1l_1}^{k_2l_2}
U^{\ast}_b(3, 1)^{s_2r_2}_{s_1 r_1} U_a(3, 1)_{s_1t_1}^{s_2t_2}
U^{\ast}_a(3, 1)^{k_2^{\prime}j_2^{\prime}}_{k_1^{\prime} j_1^{\prime}} U_b(3, 1)_{k_1^{\prime}l_1^{\prime}}^{k_2^{\prime}l_2^{\prime}}
U^{\ast}_b(3, 1)^{s_2^{\prime}r_2^{\prime}}_{s_1^{\prime} r_1^{\prime}} U_a(3, 1)_{s_1^{\prime}t_1^{\prime}}^{s_2^{\prime}t_2^{\prime}} \nonumber\\
&= \sumak{31}\lambdaa{31}
\Delta^{k_{31}^a}_{j_1l_1j_1^{\prime}l_1^{\prime}t_1r_1t_1^{\prime}r_1^{\prime}}
\Delta^{k_{31}^a}_{j_2l_2j_2^{\prime}l_2^{\prime}t_2r_2t_2^{\prime}r_2^{\prime}} \\
&\inttwo dU_a(3,2)dU_b(3,2) \nonumber\\
&\qquad\sum_{\substack{k_3,k_4\\s_3,s_4\\}}
\sum_{\substack{k^{\prime}_3,k^{\prime}_4\\s^{\prime}_3,s^{\prime}_4\\}} 
U^{\ast}_a(3, 2)^{k_4j_4}_{k_3 j_3} U_b(3, 2)_{k_3l_3}^{k_4l_4}
U^{\ast}_b(3, 1)^{s_4r_4}_{s_3 r_3} U_a(3, 2)_{s_3t_3}^{s_4t_4}
U^{\ast}_a(3, 2)^{k_4^{\prime}j_4^{\prime}}_{k_3^{\prime} j_3^{\prime}} U_b(3, 2)_{k_3^{\prime}l_3^{\prime}}^{k_4^{\prime}l_4^{\prime}}
U^{\ast}_b(3, 2)^{s_4^{\prime}r_4^{\prime}}_{s_3^{\prime} r_3^{\prime}} U_a(3, 2)_{s_3^{\prime}t_3^{\prime}}^{s_4^{\prime}t_4^{\prime}} \nonumber\\
&= \sumak{31}\lambdaa{31}
\Delta^{k_{32}^a}_{j_3l_3j_3^{\prime}l_3^{\prime}t_3r_3t_3^{\prime}r_3^{\prime}}
\Delta^{k_{32}^a}_{j_4l_4j_4^{\prime}l_4^{\prime}t_4r_4t_4^{\prime}r_4^{\prime}}
\label{int6}
\end{flalign}
hold. Substituting (\ref{int1}), (\ref{int2}), (\ref{int3}), (\ref{int4}), (\ref{int5}), and (\ref{int6}) to (\ref{example-second-frame}), we get
\begin{flalign}
\faltbase{2}{3}{n}{m} &= 
\sumak{11}\sumak{31}\sumbk{11}\sumak{12}\sumak{32}\sumbk{12}\lambdaa{11}\lambdaa{31}\lambdab{11}\lambdaa{12}\lambdaa{32}\lambdab{12} \nonumber\\
&
\inttwo dU_a(2,1)dU_b(2,1)
\left[
\sum_{\substack{u_1 u^{\prime}_1 i_1 i^{\prime}_1\\
j_1 j^{\prime}_1 l_1 l^{\prime}_1}}
\sum_{\substack{r_1 r^{\prime}_1 t_1 t^{\prime}_1\\
p_1 p^{\prime}_1 q_1 q^{\prime}_1}}
\Delta^{k_{11}^a}_{u_10u^{\prime}_10i_10i^{\prime}_10}
\Delta^{k_{31}^a}_{j_1l_1 j^{\prime}_1 l^{\prime}_1 t_1 r_1 t^{\prime}_1 r^{\prime}_1} 
\Delta^{k_{11}^b}_{p_10p^{\prime}_10q_1 0q^{\prime}_10}
\right. \\
&\qquad\left. 
U_a(2, 1)_{t_1 u_1}U_a(2, 1)_{t^{\prime}_1 u^{\prime}_1}
U_a^{\ast}(2, 1)_{j_1 i_1}U_a^{\ast}(2, 1)_{j^{\prime}_1 i^{\prime}_1}
U_b(2, 1)_{l_1 p_1}U_b(2, 1)_{l^{\prime}_1 p^{\prime}_1}
U_b^{\ast}(2, 1)_{r_1 q_1}U_b^{\ast}(2, 1)_{r^{\prime}_1 q^{\prime}_1}\right] \times \nonumber\\
&\inttwo dU_a(2,2)dU_b(2,2) \left[ 
\sum_{\substack{u_2 u^{\prime}_2 i_2 i^{\prime}_2\\
j_2 j^{\prime}_2 l_2 l^{\prime}_2}}
\sum_{\substack{r_2 r^{\prime}_2 t_2 t^{\prime}_2\\
p_2 p^{\prime}_2 q_2 q^{\prime}_2}}
\sum_{\substack{u_3 u^{\prime}_3 i_3 i^{\prime}_3\\
j_3 j^{\prime}_3 l_3 l^{\prime}_3}}
\sum_{\substack{r_3 r^{\prime}_3 t_3 t^{\prime}_3\\
p_3 p^{\prime}_3 q_3 q^{\prime}_3}}
\Delta^{k_{11}^a}_{u_20u^{\prime}_20i_20i^{\prime}_20}
\Delta^{k_{31}^a}_{j_2l_2 j^{\prime}_2 l^{\prime}_2 t_2 r_2 t^{\prime}_2 r^{\prime}_2} 
\Delta^{k_{11}^b}_{p_20p^{\prime}_2q_2 0q^{\prime}_20}
\right. \nonumber\\
&\qquad\Delta^{k_{12}^a}_{u_30u^{\prime}_30i_30i^{\prime}_30}
\Delta^{k_{32}^a}_{j_3l_3 j^{\prime}_3 l^{\prime}_3 t_3 r_3 t^{\prime}_3 r^{\prime}_3} 
\Delta^{k_{12}^b}_{p_30p^{\prime}_3q_3 0q^{\prime}_30}
U_a(2, 2)_{t_2 u_2}^{t_3 u_3}U_a(2, 2)_{t^{\prime}_2 u^{\prime}_2}^{t^{\prime}_3 u^{\prime}_3}
U_a^{\ast}(2, 2)_{j_2 i_2}^{j_3 i_3}
U_a^{\ast}(2, 2)_{j^{\prime}_2 i^{\prime}_2}^{j^{\prime}_3 i^{\prime}_3} \nonumber\\
&\qquad U_b(2, 2)_{l_2 p_2}^{l_3 p_3}
U_b(2, 2)_{l^{\prime}_2 p^{\prime}_2}^{l^{\prime}_3 p^{\prime}_3}
U_b^{\ast}(2, 2)_{r_2 q_2}^{r_3 q_3}
U_b^{\ast}(2, 2)_{r^{\prime}_2 q^{\prime}_2}^{r^{\prime}_3 q^{\prime}_3}
\left. \right]\times \nonumber\\
&
\inttwo dU_a(2,3)dU_b(2,3)
\left[
\sum_{\substack{u_4 u^{\prime}_4 i_4 i^{\prime}_4\\
j_4 j^{\prime}_4 l_4 l^{\prime}_4}}
\sum_{\substack{r_4 r^{\prime}_4 t_4 t^{\prime}_4\\
p_4 p^{\prime}_4 q_4 q^{\prime}_4}}
\Delta^{k_{12}^a}_{u_40u^{\prime}_40i_40i^{\prime}_40}
\Delta^{k_{32}^a}_{j_4l_4 j^{\prime}_4 l^{\prime}_4 t_4 r_4 t^{\prime}_4 r^{\prime}_4} 
\Delta^{k_{12}^b}_{p_40p^{\prime}_40q_4 0q^{\prime}_40}
\right. \\
&\qquad\left. 
U_a(2, 3)_{t_4 u_4}U_a(2, 3)_{t^{\prime}_4 u^{\prime}_4}
U_a^{\ast}(2, 3)_{j_4 i_4}U_a^{\ast}(2, 3)_{j^{\prime}_4 i^{\prime}_4}
U_b(2, 3)_{l_4 p_4}U_b(2, 3)_{l^{\prime}_4 p^{\prime}_4}
U_b^{\ast}(2, 3)_{r_4 q_4}U_b^{\ast}(2, 3)_{r^{\prime}_4 q^{\prime}_4}\right] \allowdisplaybreaks\nonumber \\
&= \sumak{11}\sumak{31}\sumbk{11}\sumak{12}\sumak{32}\sumbk{12}\lambdaa{11}\lambdaa{31}\lambdab{11}\lambdaa{12}\lambdaa{32}\lambdab{12} \nonumber\\ &\qquad\inttwo  dU_a(2,1)dU_b(2,1)\delthree{1} \times \nonumber\\
&\qquad\inttwo dU_a(2,2)dU_b(2,2) \delsix{1}{2} \times\nonumber\\
&\qquad\inttwo dU_a(2,3)dU_b(2,3)\delthreebase{2}{3},
\end{flalign}
which is the right hand side of the first equality in (\ref{integrate-blocks}) when $n/m = 2$.
\subsection{Proof of Theorem \ref{second-moment-alt}}

Similar to Theorem \ref{second-moment-2}, we only show the inequality for $\faltcustom{2}{3}$ 
here. The inequality for $\faltcustom{2}{2}$ can be shown in the same manner. In the process of showing the final inequality of the theorem, we expand $\athreem$ and $B(3, m)$ as the sum of a vector/matrix whose components are $O(1)$ and a vector/matrix whose components are $O(1/2^{m/2})$.

To evaluate Eq.~(\ref{a3m})\footnote{The evaluation procedure is straightforward, but a lot of computation is required. Thus, instead of performing a hand-calculation, we built an algorithm to evaluate (\ref{a3m}) for arbitrary $m$ and derived the expansion formula by computational calculation. We also built an algorithm for evaluating (\ref{b3m}) and derived the expansion formula by computational calculation.}, we can expand $\athreem$ as
\begin{flalign}
	\athreem = \frac{1}{2^{m}}\left(\vzero + \frac{1.2}{2^{m/2}}\vone \right), 
\end{flalign}
where 
\begin{flalign}
\vzeroi &= 
\begin{cases}
1 & \quad i=1 \quad(k_a, k_b, k_c=1), \\
1 & \quad i=22 \quad(k_a,k_b,k_c=2), \\
0 & \quad \mbox{otherwise}, 
\end{cases}
\qquad
\qquad |\vonei | < 1 . 
\end{flalign}
Also, evaluating Eq.~(\ref{b3m}), we can expand $B(3, m)$ as
\begin{flalign}
B(3, m) = \frac{1}{2^{2m}}\left(D + \frac{1.3}{2^{m/2-6}}X\right),
\end{flalign}
where 
\begin{flalign}
D_{ij} &=
\begin{cases}
1 & \quad i=1, j=1 \quad(k_a, k_b, k_c, k_d, k_e, k_f=1), \\
1 & \quad i=22, j=22 \quad(k_a, k_b, k_c, k_d, k_e, k_f=2), \\
0 & \quad \mbox{otherwise}, 
\end{cases} 
\qquad
\qquad |X_{ij}| < \frac{1}{64}.
\end{flalign}
With $\alpha = n/m$, let $\gxd{k}{\alpha}$ be as the set of matrices expressed by $\prod_{i=1}^{\alpha}R_i$ where $R_i = D {\rm \ or \ } X$ and the number of $X$s in $\{R_i\}$ is $k$. For example, $XDXX \in \gxd{3}{4}$ and $XDDD \in \gxd{1}{4}$. Then, $\faltcustom{2}{3}$ is expanded as
\begin{flalign}
	\faltcustom{2}{3} &=  \frac{1}{2^{2m}}\left(\vzero^T + \frac{1.2}{2^{m/2}}\vone^T \right) 
	 \frac{1}{2^{2n-2m}}\left(D+\frac{1.3}{2^{m/2-6}}X\right)^{\alpha-1}
	 \left(\vzero + \frac{1.2}{2^{m/2}}\vone \right) \nonumber\\
	 &= \frac{1}{2^{2n}} \left(\vzero^T D^{\alpha-1}\vzero + 
	 \left(\frac{2.4}{2^{m/2}}\right)\vone^T D^{\alpha-1}\vzero +
	 	 \left(\frac{1.2^2}{2^{m}}\right)\vone^T D^{\alpha-1}\vone \right) \nonumber\\
	 &+ \frac{1}{2^{2n}}\sum_{k=1}^{\alpha - 1}\left(\frac{1.3}{2^{m/2-6}}\right)^k 
	 \sum_{i=1}^{{}_{\alpha - 1} \mathrm{C} _k}
	 \left(\vzero^T g^k_{\alpha i} \vzero + \left(\frac{2.4}{2^{m/2}}\right)\vone^T g^k_{\alpha i} \vzero +\left(\frac{1.2^2}{2^{m}}\right)\vone^T g^k_{\alpha i} \vone
	 \right),
\end{flalign}
where $g^k_{\alpha i}$ ($i=1,2\dots {}_{\alpha} \mathrm{C} _k$) is an element of $\gxd{k}{\alpha}$.
For an arbitrary $g\in\gxddef$ with $k\geq 1$, 
\begin{flalign}
\label{g-inequality}
\vvec{r}^{\rm T} g \vvec{s} &< (1,1,\cdots,1)
\begin{pmatrix}
1 & 0 \cdots & 0\\
0 & 1 \cdots & 0\\
\vdots & \vdots & \vdots \\
0 & 0 \cdots & 1
\end{pmatrix}^{\alpha - k}
\begin{pmatrix}
\frac{1}{64} & \frac{1}{64} \cdots & \frac{1}{64}\\
\frac{1}{64} & \frac{1}{64} \cdots & \frac{1}{64}\\
\vdots & \vdots & \vdots \\
\frac{1}{64} & \frac{1}{64} \cdots & \frac{1}{64}
\end{pmatrix}^{k}
\begin{pmatrix}
1\\
1\\
\vdots\\
1
\end{pmatrix} = 64
\end{flalign}
holds. For $D$ and $\vvec{0}, \vvec{1}$
\begin{flalign}
\label{vector-inequality}
\vvec{1}^{\rm T} D^{\frac{n}{m}-1} \vvec{s} &< \sum^{64}_{i=1}\left(1 \cdot  (\delta_{i1} + \delta_{i22}) \cdot 1\right) = 2 \\
\label{vector-inequality-2}
\vvec{0}^{\rm T} D^{\frac{n}{m}-1} \vvec{0} &= 2
\end{flalign}
holds where $r,s=0,1$. By using the inequalities (\ref{g-inequality}), (\ref{vector-inequality}), and (\ref{vector-inequality-2}), the upper bound for $\faltcustom{2}{3}$
is derived as follows:
\begin{flalign} 
\label{proof-f-inequality}
	\faltcustom{2}{3} &<  
	 \frac{1}{2^{2n}} \left(2 + 2\left(\frac{2.4}{2^{m/2}}\right) +
	 	 2\left(\frac{1.2^2}{2^{m}}\right) + 64\sum_{k=1}^{\alpha-1} {}_{\alpha-1} \mathrm{C} _k\left(\frac{1.3}{2^{m/2-6}}\right)^k 1^k
	 	 \left(1+\frac{1.2}{2^{m/2}} \right)^2
	 	  \right)  \nonumber\\ 
	 &= \frac{1}{2^{2n-1}}\left(1+\frac{1.2}{2^m}\right)^2\left(1+ 32\left(\left(1 + \frac{1.3}{2^{m/2-6}}\right)^{\alpha-1} -1\right)\right)\nonumber\\
	  &= \left(1+\frac{1}{2^{n}}\right)\left(1+\frac{1.2}{2^m}\right)^2\left(1+ 32\left(\left(1 + \frac{83.2}{2^{m/2}}\right)^{\alpha-1} -1\right)\right)\fhaarn{2}.
\end{flalign}

\subsection{Proof of Corollary \ref{large-n}}
As in the above theorems, we only show the inequality for $\faltcustom{2}{3}$ here. When $m = 2a \log_2 n$, 
\begin{flalign}
\label{proof-exponential}
	\left(1 + \frac{83.2}{2^{m/2}}\right)^{\alpha-1} <  \left[\left(1 + \frac{83.2}{2^{m/2}}\right)^{\frac{2^{m/2}}{83.2}}\right]^{\frac{83.2n}{2^{m/2}m}} < e^{\frac{83.2n}{2^{m/2}m}} = e^{\frac{41.6}{a n^{a-1}\log_2 n}}.
\end{flalign}
If $41.6/(a n^{a-1}\log_2 n) < 1$, 
\begin{flalign}
\label{proof-exponential-2}
	e^{\frac{41.6}{a n^{a-1}\log_2 n}} < 1+(e-1)\frac{41.6}{a n^{a-1}\log_2 n}.
\end{flalign}
Substituting Eqs.~(\ref{proof-exponential}), (\ref{proof-exponential-2}), and 
$m = 2a \log_2 n$ into (\ref{proof-f-inequality}), we get
\begin{flalign}
\faltcustom{2}{3} &<  
	\left(1+\frac{1}{2^{n}}\right)\left(1+\frac{1.2}{n^{2a}}\right)^2\left(1+ \frac{2288}{an^{a-1}\log_2 n}\right)\fhaarn{2}.
\end{flalign}

\bibliographystyle{main}
\bibliography{main}

\end{document}